\documentclass{emulateapj} 
\usepackage{amsmath,amssymb,amsfonts} 
\usepackage{natbib}
\def\cm{{\rm\,cm}}

\def\gm{{\rm\,g}}

\def\au{{\rm AU}}

\def\AU{{\rm\, AU}}

\def\K{{\rm\,K}}  
\def\yr{{\rm\,yr}}

\def\ts{t_{\rm stop}}

\def\Hg{H_{\rm g}}

\def\OmegaK{\Omega_{\rm K}}

\def\taus{\tau_{\rm s}}

\def\rhos{\rho_{\rm s}}
\def\rhog{\rho_{\rm g}}

\def\rhod{\rho_{\rm d}}

\def\Sigmad{\Sigma_{\rm d}}
\def\Sigmag{\Sigma_{\rm g}}

\def\cs{c_{\rm s}}

\def\lesssim{\mathrel{\hbox{\rlap{\hbox{\lower4pt\hbox{$\sim$}}}\hbox{$<$}}}}
\def\gtrsim{\mathrel{\hbox{\rlap{\hbox{\lower4pt\hbox{$\sim$}}}\hbox{$>$}}}}

\def\Zr{Z_{\rm rel}}

\def\vmax{v_{\rm max}}
\def\zmax{z_{\rm max}}

\def\muinit{\mu_{\rm init}}

\def\vrel{v_{\rm rel}}
\def\zhalf{z_{\rm 50}}
\def\zw{z_{\rm w}}
\def\Ricrit{Ri_{\rm crit}}

\newcommand{\p}{\partial}

\begin{document}

\title{Forming Planetesimals by Gravitational Instability \\II. How Dust Settles to its Marginally Stable State}

\author{Aaron T. Lee\altaffilmark{1}, Eugene Chiang\altaffilmark{1,2}, Xylar Asay-Davis\altaffilmark{3}, Joseph Barranco\altaffilmark{4}}

\altaffiltext{1}{Department of Astronomy, University of California Berkeley,
    Berkeley, CA 94720}
\altaffiltext{2}{Department of Earth and Planetary Science, University of California Berkeley,
    Berkeley, CA 94720}
\altaffiltext{3}{Center for Nonlinear Studies, Los Alamos National Laboratory, Los Alamos, NM 87545}    
\altaffiltext{4}{Department of Physics and Astronomy, San Francisco State University,
    San Francisco, CA 94132}
    
\email{a.t.lee@berkeley.edu}

\begin{abstract}

  Dust at the midplane of a circumstellar disk can become
  gravitationally unstable and fragment into planetesimals if the
  local dust-to-gas ratio $\mu_0\equiv \rhod/\rhog$ is sufficiently high.   
  We simulate
  how dust settles in passive disks and ask how high $\mu_0$ can
  become. We implement a hybrid scheme that alternates between a 1D
  code to settle dust and a 3D shearing box code to test for dynamical
  stability.  This scheme allows us to explore the behavior of small
  particles having short but non-zero stopping times in gas: $0 < t_{\rm
    stop} \ll$ the orbital period. The streaming instability is thereby filtered out. Dust settles until Kelvin-Helmholtz-type instabilities at the top and bottom faces of the dust layer threaten to overturn the entire layer.  In
  this state of marginal stability, $\mu_0 = 2.9$ for a disk whose
  bulk (height-integrated) metallicity $\Sigmad/\Sigmag$ is
  solar---thus $\mu_0$ increases by more than two orders of magnitude
  from its well-mixed initial value of $\mu_{\rm 0,init} =
  \Sigmad/\Sigmag = 0.015$.  For a disk whose bulk metallicity is
  $4\times$ solar ($\mu_{\rm 0,init} = \Sigmad/\Sigmag = 0.06$), the
  marginally stable state has $\mu_0 = 26.4$.  These maximum values of
  $\mu_0$, which depend on the background radial pressure gradient,
  are so large that gravitational instability of small
  particles is viable in disks whose bulk metallicities are just a few
  ($\lesssim 4$) times solar. Our result supports earlier studies that
  assumed that dust settles until the Richardson number $Ri$ is
  spatially constant.  Our simulations are free of this assumption but
  provide evidence for it within the boundaries of the dust layer,
  with the proviso that $Ri$ increases with $\Sigmad/\Sigmag$ in the
  same way that we found in Paper I.  Because increasing the dust
  content decreases the vertical shear and increases stability, the
  midplane $\mu_0$ increases with $\Sigmad/\Sigmag$ in a faster than
  linear way, so fast that modest enhancements in $\Sigmad/\Sigmag$
  can spawn planetesimals directly from small particles.
  \end{abstract}

\section{INTRODUCTION}
\label{sec:introduction}

Dust can settle quickly in gaseous protoplanetary disks. 
In a passive (non-turbulent) nebula, a particle's vertical height $z$
above the midplane obeys
\begin{equation}
\ddot{z} = -\dot{z}/\ts - \OmegaK^2 z
\end{equation}
where the first term on the right-hand side accounts for gas drag, and
the second term accounts for stellar gravity when $z \ll r$, the cylindrical radius. Here $\OmegaK$ is the Keplerian orbital frequency
and 
\begin{equation} \label{eqn:tstop}
\ts \equiv \frac{m v_{\rm rel}}{F_{\rm D}} 
\end{equation}
is the momentum stopping time of a particle of mass $m$ moving at
speed $v_{\rm rel}$ relative to gas. Expressions for the drag force
$F_{\rm D}$ can be found in \citet{adachietal76} and
\citet{weidenschilling77}.  We are interested in small, well-coupled
particles having non-zero stopping times much shorter than the dynamical time:
$0 < \taus \equiv \OmegaK \ts \ll 1$. Spherical particles of radius $s$
and internal density $\rhos$ that experience Epstein drag 
($F_{\rm D} \propto s^2 \vrel$ so that $t_{\rm stop}$ does not depend on $\vrel$) 
settle to the midplane at terminal velocity
$-\OmegaK^2z \ts$ in a time
\begin{equation} \label{eq_tz} 
t_{\rm settle} \sim \frac{1}{\OmegaK \taus} \sim 10^3 \left( \frac{0.1 \cm}{s} \right) \left( \frac{1 \gm \cm^{-3}}{\rhos} \right) \left( \frac{F}{1} \right) \yr \,.
\end{equation}
For this and all other numerical evaluations in this paper, we use a
background disk that is $F$ times more massive than the minimum-mass
nebula of Chiang \& Youdin (2010, hereafter CY10; see Appendix
\ref{app:background}).  For such a disk $t_{\rm settle}$ is
nearly independent of stellocentric distance. The assumption that particles
are spherical may not be too bad because fractal aggregates of
grains are expected to compactify as they collide with one another
\citep{dominiktielens97, dullemonddominik05, ormeletal07}.

For millimeter-sized particles, the settling time $t_{\rm settle}$ is
much shorter than the disk lifetime, measured in Myr
\citep{hillenbrand05, hernandezetal08}. By comparison, micron-sized and
smaller particles stay suspended at least one scale height above the
midplane as long as the gas disk is present.  To the extent that
collections of particles of different sizes tend to place their mass
at the upper end of the size distribution and their surface area at
the lower end, we can expect most of the solid mass in disks to
sediment out into a thin sublayer, leaving behind the smallest of grains to
absorb incident starlight in a flared disk atmosphere.
On the whole this picture is consistent
with observed spectral energy distributions of T Tauri disks,
although some models hint that large grains might remain lofted up in
a disk two gas scale heights thick \citep{dalessioetal06}.  Settling
can only proceed when and where disk turbulence dies, in regions
where gas is insufficiently dense to sustain gravitoturbulence \citep{gammie01} and
too poorly ionized to be magnetorotationally unstable
\citep{gammie96}. Our current understanding of disk turbulence
easily admits such passive regions. \citet{turnercarballidosano10}
found in numerical simulations that even when disk surface layers were magnetorotationally unstable,
grains at the midplane settled much as they would in a laminar flow.
Recently \citet{perezbeckerchiang10} estimated that practically the entire
disk would be immune to the magnetorotational instability because ion
densities are too low for the plasma to drive turbulence in the
overwhelmingly neutral gas.

In passive disk regions, how far down do dust particles settle?  
When particles are small enough not to be affected by aerodynamic
streaming instabilities (e.g., \citealt{baistone10}), we expect them to 
settle until the dust density gradient $\partial
\rhod/\partial z$ becomes so large, and the consequent vertical shear
in orbital velocity $\partial v_\phi/\partial z$ so strong, that the
sublayer is on the verge of overturning by a Kelvin-Helmholtz-type
instability \citep[KHI;][]{weidenschilling80}. An order-of-magnitude
estimate of the minimum layer thickness can be derived using the
Richardson number
\begin{equation}
\label{eqn:richardson}
	Ri \equiv \frac{-(g/\rho)(d\rho/dz)}{(dv_\phi/dz)^2}  \,,
\end{equation}
which if less than some critical value $Ri_{\rm crit}$ may signal that
the layer is KH unstable \citep[e.g.,][]{drazinreid04}.  Here $g$ is
the vertical gravitational acceleration and $\rho = \rhod + \rhog$ is
the total density of dust plus gas.  In Lee et al. (2010, hereafter
Paper~I), we found that $Ri_{\rm crit}$ increases with $\Sigmad/\Sigmag$,
the ratio of dust to gas surface densities, a.k.a.~the bulk
metallicity.  For disks of bulk solar metallicity, we determined empirically
that $Ri_{\rm crit} \approx 0.2$.

To translate the Richardson number (\ref{eqn:richardson}) into a
critical dust layer thickness, first recognize that the orbital
velocity $v_\phi$ depends on the local dust-to-gas ratio $\mu \equiv
\rhod/\rhog$ according to
\begin{equation} \label{eqn:vphi}
v_\phi = \OmegaK r \left( 1 - \frac{\eta}{\mu+1} \right)
\end{equation}
in the inertial frame, where
\begin{eqnarray}
\label{eqn:eta}
\eta & \equiv &
 \frac{ -(1/\rho_{\rm g}) \partial P / \partial r }{2 \Omega_{\rm K}^2r} \nonumber \\
 & \approx & \frac{1}{2} \left( \frac{c_{\rm s}}{\OmegaK r} \right)^2 \approx \frac{1}{2} \left( \frac{H_{\rm g}}{r} \right)^2 \nonumber \\
 & \approx & 8 \times 10^{-4} \left( \frac{r}{\rm AU} \right)^{4/7} 
\end{eqnarray}
is a dimensionless measure of the strength of the background radial
pressure gradient $\partial P/\partial r$, with gas scale height
$H_{\rm g}$ and sound speed $c_{\rm s}$ (e.g., \citealt{nakagawaetal86}).
When $\partial P/\partial r < 0$,
pressure provides extra support against radial stellar gravity
and so drives the gas to move on
slower than Keplerian orbits. The orbital velocity depends on $\mu$ as in
(\ref{eqn:vphi}) because dust-laden gas, weighed down by the
extra inertia of solids, is accelerated less by the radial pressure
gradient than is dust-free gas, and so must hew more closely to
Keplerian rotation.  Call the critical
layer height $\Delta z_{\rm Ri}$ for which $Ri = Ri_{\rm
  crit}$, and assume the midplane gas-to-dust ratio $\mu_0 \gtrsim 1$
(above the layer $\mu \ll 1$).
Then evaluating equation (\ref{eqn:richardson}) with
the approximations $g \approx -\OmegaK^2 \Delta z_{\rm Ri}$ (no self-gravity),
$\rho^{-1} \partial \rho/ \partial z \sim -1 /\Delta z_{\rm
  Ri}$, and $\partial v_{\phi}/\partial z \sim -\eta \OmegaK r / \Delta z_{\rm Ri}$, we find
\begin{eqnarray}
\Delta z_{\rm Ri} & \sim & Ri_{\rm crit}^{1/2} \eta r \nonumber \\
 & \sim & \frac{1}{2} Ri_{\rm crit}^{1/2} \frac{H_{\rm g}}{r} H_{\rm g} \nonumber \\
 & \sim & 5 \times 10^{-3} \left( \frac{Ri_{\rm crit}}{0.2} \right)^{1/2} \left( \frac{r}{\rm AU} \right)^{2/7} H_{\rm g} \,.
\label{eqn:deltaz}
\end{eqnarray}

Equation (\ref{eqn:deltaz}) indicates the dust layer could be quite
thin, subtending on the order of 1\% of the gas scale height. Is this
thin enough for the dust to self-gravitate and hopefully
fragment into planetesimals?
One can compare the
midplane density to the ``Toomre density'' required for the disk to
undergo gravitational instability on the dynamical time
$\OmegaK^{-1}$ \citep{safronov69,goldreichward73}.\footnote{Even if
  the Toomre density is attained so that marginally bound clumps of dust
  and gas form in $\OmegaK^{-1}$ time, continued collapse of dust is
  not guaranteed. For dust to concentrate further it must sediment to
  the centers of the gas clumps over timescales $t_{\rm settle} \gg
  \OmegaK^{-1}$. During this time the dust clumps are susceptible to
  erosion by gas streaming or turbulence
  \citep[e.g.,][]{cuzzietal08}. \label{foot:slow} }
As reviewed by
CY10, the Toomre density is\footnote{Strictly speaking, the Toomre
  criterion for gravitational instability is derived for
  two-dimensional disks characterized by surface densities, not volume
  densities \citep{toomre64,goldreichbell65}. To derive
  our Toomre volume density, we assign a half-thickness to the disk
  equal to $c/\OmegaK$, where $c$ is the velocity dispersion of the
  dust + gas mixture. This assignment is not rigorous; see CY10.}
\begin{equation} \label{eqn:Toomrerho}
\rho_{\rm Toomre} \approx \frac{M_{\ast}}{2\pi r^3} \approx 10^{-7} \left( \frac{r}{\au} \right)^{-3} \gm \cm^{-3} 
\end{equation}
where the numerical evaluation is for a central stellar mass $M_{\ast}$ equal to $1 M_{\odot}$.
Now the actual midplane (subscript 0) density is
\begin{equation}
\rho_0 = \rho_{\rm g0} + \rho_{\rm d0} = 2.7 \times 10^{-9} F (1 + \mu_0) \left( \frac{r}{\rm AU} \right)^{-39/14} \gm \cm^{-3}
\end{equation}
which means the midplane dust-to-gas ratio must be
\begin{equation} \label{eqn:muToomre}
\mu_{\rm 0, Toomre} \approx 34 \left( \frac{1}{F} \right) \left( \frac{M_{\ast}}{M_\odot} \right) \left( \frac{r}{\rm AU} \right)^{-3/14}
\end{equation}
for the midplane density to match the Toomre density. By comparison,
in our crude model of a dust sublayer whose height above the midplane
cannot be
smaller than $\Delta z_{\rm Ri}$, the midplane dust-to-gas ratio
cannot exceed
\begin{equation} \label{eqn:muRi}
\mu_{\rm 0,Ri} \sim \frac{ \Sigmad/ (2\Delta z_{\rm Ri}) }{\rho_{\rm g0}} \sim 1 \left( \frac{\Sigmad/\Sigmag}{0.015} \right) \left( \frac{0.2}{Ri_{\rm crit}} \right)^{1/2} \left( \frac{r}{\rm AU} \right)^{-4/7}
\end{equation}
which is nominally smaller than $\mu_{\rm 0,Toomre}$ by more than an
order of magnitude. Here the bulk (height-integrated) metallicity
$\Sigmad/\Sigmag$ is normalized to solar abundance \citep{lodders03},
assuming all metals have condensed into
grains.\footnote{ \label{foot:metallicity} The assumption that all
  metals have condensed is valid only for temperatures $T \la 41$
  K. For 41 K $\la T \la$ 182 K, methane ice sublimates and
  $\Sigmad/\Sigmag \approx 0.78 \times 0.015$ (\citealt{lodders03};
  CY10). For our fiducial disk described in Appendix \ref{app:background},
  $T \la 182$ K for $r \ga 0.4$ AU.
For $T \ga 182$ K, water and other ices sublimate and
  the maximum $\Sigmad/\Sigmag \approx 0.33 \times 0.015$. Reductions
  in $\Sigmad/\Sigmag$ due to sublimation may be offset by radial pileups
  of dust.}

For many years the fact that $\mu_{\rm 0,Ri}$ falls short of $\mu_{\rm
  0,Toomre}$ was believed to rule out the formation of planetesimals
by collective effects, self-gravitational or otherwise
\citep[e.g.,][]{weidenschillingcuzzi93}.  But there are more ways to
achieve the Toomre density than vertical settling.  A dissipative form
of gravitational instability can, in principle, collect particles
radially into overdense rings even when self-gravity is weaker than
stellar tidal forces (\citealt{ward76}; \citealt{ward00};
\citealt{coradinietal81}; \citealt{youdin05a}; for a simple explanation
of the instability, see the introduction of
\citealt{goodmanpindor00}).  It is not clear whether this instability,
which operates over lengthscales and timescales longer than those
characterizing the Toomre instability by at least a factor of
$(\mu_{\rm 0,Toomre}/\mu_0)^2$, can compete with other effects that
seek to rearrange dust and gas \citep[e.g.,][]{youdin05a}.

Another alternative is to invoke larger dust particles that are
only marginally coupled to gas; these can clump by the aerodynamic
streaming instability (SI; 
\citealt{youdingoodman05}; \citealt{johansenetal09};
\citealt{baistone10}; \citealt{baistone10b}).  In their 3D numerical
simulations, \citet{johansenetal09} reported that particles having
$\taus = 0.1$--0.4---corresponding to sizes of a few centimeters at
$r= 5$ AU if $F=1$, and larger sizes if $F>1$---concentrated so
strongly by aerodynamic effects that planetesimals effectively
hundreds of kilometers across coalesced within just a few orbits. To
obtain this result, \citet{johansenetal09} initialized their
simulations by placing the bulk of the disk's solid mass into particles
approaching decimeters in size. \citet{baistone10} greatly expanded
the range of $\taus$ modeled and found similar results for their 3D
simulations: in the highly turbulent states driven by the SI,
instantaneous densities exceeded the Roche density\footnote{The Roche
  density is that required for a fluid satellite to be gravitationally
  bound against tidal forces exerted by a central body. It is greater
  than the Toomre density by a factor of $\sim$$7\pi$, and is the more
  appropriate threshold density for the highly localized clumps of dust
  created by the SI.} when the
disk's solids were all composed of particles having $\taus = 0.1$--1
and the bulk height-integrated metallicity was about twice solar; see
run R10Z3-3D in their Figure 5.  For this same run, the time-averaged
dust-to-gas ratio at the midplane was $\sim$12, a factor of a few less
than the Toomre threshold; see their Figure 4 and compare with our
equation (\ref{eqn:muToomre}). By contrast, when half or more of the
disk's solid mass had $\taus < 0.1$, or when disks had smaller
metallicities, their simulated densities fell short of the Roche and
Toomre densities by more than an order of magnitude.  Note that we are
quoting from the 3D simulations of \citet{baistone10}.

Given how sensitive the SI is to the existence of marginally coupled
particles (centimeter to meter sized for $\taus \sim 0.1$--1, $r \sim
1$--30 AU, and order unity $F$), whether enough such particles
actually exist in protoplanetary disks for the SI to play a dominant
role in planetesimal formation remains an open and delicate
issue. Appeal is often made to observed spectral energy distributions
and images of T Tauri disks at centimeter wavelengths; these suggest
that much of the solid mass is in millimeter to centimeter sized
particles \citep[e.g.,][]{dalessioetal01,wilneretal05}.  Larger sized
particles are plausibly also present but are not inferred for want of
data probing the disk at longer wavelengths.
One problem concerns how quickly $\taus \gtrsim 0.1$ particles can be
grown, and how they can survive orbital decay by gas drag.  
In the 3D simulations of \citet{baistone10},
the SI clumped particles strongly enough for self-gravity to be significant
when $\taus \gtrsim 0.1$ particles
comprised more than half of the disk's solid mass.
It is unclear whether
particle-particle sticking can build up such a population before it is
lost to the star by gas drag. This concern is ameliorated by
enhancements in particle density (pileups) that may occur as particles
drift radially inward \citep{youdinshu02,youdinchiang04}, and by the
reduction of drift speeds brought about by multiple particle sizes
(\citealt{baistone10}; see their Figure 8).

Regardless of which scenario nature prefers---particle concentration
by the streaming instability; dissipative gravitation into rings; or
dynamical collapse of a vertically settled sublayer, which is the subject of
this paper---all the proposed ways of forming planetesimals depend on
knowing how far down dust settles and what maximum dust-to-gas ratios
$\mu_0$ can be attained at the midplane. Our order-of-magnitude
estimate in equation (\ref{eqn:muRi}) requires testing.  Among the
most realistic simulations of particle settling
are those by \citet{johansenetal09} and \citet{baistone10}, both of
which concentrated on the SI.  \citet{johansenetal09} reported that
super-centimeter sized particles settled into sublayers in which the
midplane-averaged $\mu_0$ ranged from 0.6 to 9.0 as the bulk
metallicity ranged from 1 to $3$$\times$ solar.  \citet{baistone10}
found that the highest $\taus$ particles settled the most, driving
turbulence that lofted smaller $\taus$ particles to greater heights.
They argued that in their simulations, all of which were characterized
by $\max \taus \geq 0.1$, particles were so strongly stirred by the SI that the
KHI never manifested.

To complement these studies, we would like to understand the settled
equilibrium states of disks composed entirely of small particles, well
but not perfectly coupled to gas ($0< \max \taus \ll 1$),
isolated from the complicating
effects of the streaming instability but not other instabilities like
the KHI.  
Some previous attempts in this regard relied on assumed forms for
the density profile of settled dust. \citet{barranco09} presumed the
dust density profile was Gaussian in shape, and did not seek to
determine the maximum value of $\mu_0$ {\it per se}.  \citet{chiang08}
and Paper~I, following \citet{sekiya98} and \citet{youdinshu02},
assumed the dust density profile had a spatially constant Richardson
number. We found in Paper~I that under this assumption, in a disk of
bulk solar metallicity ($\Sigmad/\Sigmag = 0.015$), the sublayer could
remain KH stable for $\mu_0$ as high as 8---a value that is nearly an
order of magnitude higher than our crude estimate in (\ref{eqn:muRi}),
and as such lowers the hurdle to forming planetesimals by
gravitational instability.

\citet{weiden06} and \citet{weiden10} did not assume a shape for the
marginally stable density profile to which dust relaxes. Instead the
profile was calculated from a one-dimensional model that balanced the
downward flux of particles by gravity (including vertical
self-gravity) with the upward flux due to turbulent diffusion.  The
model used prescriptions for KH turbulence, with a
number of parameters chosen to match the 3D calculations of
\citet{cuzzietal93}---which also relied on a prescribed form of the
turbulence. \citet{weiden06} found that the density of
millimeter-sized particles could exceed the Toomre density at $r = 3$
AU in a disk approximately $1.3\times$ as massive as our minimum-mass solar nebula, and whose bulk metallicity $\Sigmad/\Sigmag = 0.054
\approx 3.6 \times$ solar.  See his Figure 10, but note that $3.6 \times$
solar metallicity $\approx 16 \times$ his ``nominal'' abundance of
solids, and that his ``critical'' density is $3 \times$ larger than
our Toomre density.

Like \citet{weiden06,weiden10}, we seek the marginally stable
state to which dust settles, in the limit that particles
are well but not perfectly coupled to gas ($0 < \taus \ll 1$).
We improve upon these earlier studies by not prescribing or parameterizing
the turbulence, but by letting turbulence arise and evolve
naturally from our 3D integrations of the standard fluid equations.
In particular, our work is free of the popular but untested assumption that
dust settles until the Richardson number equals a constant everywhere.
We allow dust grains to fall until they are stopped by whatever
instabilities they self-generate.  In the calculations presented here
we assume that dust begins well mixed with gas in a Gaussian density
profile, and then follow the dust into whatever non-Gaussian
distribution it seeks to relax.  Although we try only a Gaussian
initial profile, our method accommodates arbitrary initial conditions.

At the heart of our approach lie two codes. The first code is in one
dimension ($z$) and computes the vertical drift of dust grains at
their terminal velocities.  Though incapable of deciding whether the
density profiles it generates are prone to the KHI (or any other
instability), the 1D code can evolve dust profiles for the entire
settling time $t_{\rm settle}$, which can be arbitrarily long for
arbitrarily small grains. The task of assessing stability is reserved
for the second code: the spectral, anelastic, shearing box code of
\citet{barranco09} which treats gas and dust in the perfectly coupled
$\taus = 0$ limit. Though incapable of allowing dust to settle out of
gas, the 3D code accounts for the complicated interplay of vertical
shearing and rotational effects to decide whether a given dust layer
overturns from the KHI (or some other instability).  It tests
dynamical stability by running for dozens of dynamical times $t_{\rm
  dyn} = \OmegaK^{-1}$.  Our procedure involves alternating between
these two codes: allowing dust to settle over some fraction of the
settling timescale $t_{\rm settle}$ using the 1D code; passing the
results of the 1D code to the 3D code and allowing the dust profile to
relax dynamically over timescales $t_{\rm dyn}$; passing the results
of the 3D code back to the 1D code for further sedimentation on the
settling timescale; and so on, back and forth, until the midplane
dust-to-gas ratio stops increasing, at which point the marginally stable
state is identified.

In \S\ref{sec:method} we describe our method in full.
Results are presented in \S\ref{sec:results}, extended in \S\ref{sec:ext},
and summarized and discussed in \S\ref{sec:dissum}.

\vspace{0.2in}

\section{METHOD}
\label{sec:method}

As sketched in \S\ref{sec:introduction}, to find the marginally
stable state to which small dust grains relax, we alternate between
two codes: a 1D code that tracks how dust drifts toward the
midplane on the settling timescale $t_{\rm settle}$, and a 3D shearing
box code developed by \citet{barranco09} that allows dusty gas to
stabilize on the dynamical timescale $t_{\rm dyn} \ll t_{\rm
  settle}$. The 3D code integrates the anelastic fluid equations for
perfectly coupled dust and gas using a spectral method.  It includes a
background radial pressure gradient to drive a vertical shear.
Details about the 3D code are in \citet{barrancomarcus06},
\citet{barranco09}, and Paper~I.


Dust and gas are initially well mixed with a spatially constant density ratio:
$[\rhod(z)/\rhog(z)]_{\rm init} \equiv \muinit = {\rm constant}$. We
set $\muinit$ equal to either solar metallicity ($\muinit = 0.015$;
\citealt{lodders03}; see also footnote \ref{foot:metallicity}) or four
times solar metallicity ($\muinit = 0.06$).  To determine the initial
form of the dust density profile $\rhod(z)$, we solve the equation for
vertical hydrostatic equilibrium where gas is assumed to be initially
isothermal:
			\begin{equation} \label{eqn:vhse}
		\frac{\cs^2}{\rhog+\rhod}\frac{\p \rhog}{\p z} = -\Omega^2_{\rm K}z \, ,
		 	\end{equation}
whence
		\begin{equation} \label{eqn:rhodinit}
		\rhod = \mu_{\rm init}\rho_{\rm g0} \exp\left[-\frac{(1+\mu_{\rm init})z^2}{2H^2_{\rm g}}\right]
		\end{equation}
for constants $\muinit$, a characteristic initial height
$\Hg \equiv \cs/\OmegaK$, and the midplane gas density $\rho_{\rm g0}$. 

Taking $\muinit$ to be constant over all gas scale heights is a
simplifying but probably unrealistic assumption. Even if particles of
fixed size $s$ carry the bulk of the disk's solid mass, such
particles would not likely begin well mixed with gas
everywhere. Densities at altitude may be too low to permit particles
of size $s$ to coagulate, and to keep such particles aloft once
formed.  Nevertheless the error accrued by our assumption of constant
$\muinit$ is small insofar as increasingly small amounts of mass are
contained at larger altitudes. Moreover, we will find evidence that
the final marginally stable state to which dust relaxes is
insensitive to initial conditions (\S\ref{ssec:howconstant}). In any case, we will point out in
\S\ref{sec:results} which features of the evolving dust profile are
artifacts of our assumed initial condition.
                

Equation (\ref{eqn:rhodinit}) defines the initial dust profile used by
the 1D code, whose grid extends from $z = 0$ to $z = 3\Hg$.
The 1D Lagrangian code uses particles to track the motion of dust mass.  Each
particle represents the same amount of dust mass. Any dust density
profile $\rhod(z)$ can be converted into particle positions and back
again. The closer particles are spaced, the greater is $\rhod$.

Starting with equation (\ref{eqn:rhodinit}), we proceed as follows:

\begin{enumerate}
\item {\bf 1D code: Initialize positions of dust particles 
           and establish hydrostatic equilibrium for the gas.}

	Given $\rhod(z)$, calculate the positions of $\sim$60,000 particles
in the 1D code. Also determine the hydrostatic gas
  density $\rhog(z)$ at each particle's position by solving equation
  (\ref{eqn:vhse}).\footnote{Actually the calculation of $\rhog(z)$ can be neglected to
    good approximation, as the hydrostatic gas density deviates only
    slightly from a Gaussian throughout the evolution. Even when $\mu \gg 1$
    near the midplane, $\Delta \rhog / \rhog \sim (z/H_{\rm g})^2
    \mu$, which for our parameters remains much less than unity. In fact,
    the gas density is practically constant 
    once the dust falls to $z \lesssim 0.1 H_{\rm g}$.\label{foot:muchado}}

\item {\bf 1D code: Settle dust particles by one timestep
	 $\Delta t$.}
	
  By equating the vertical gravitational force $\propto z$ to the
  Epstein drag force $F_{\rm D} \propto \rhog v_{\rm rel}$ \citep[e.g.,][]{weidenschilling77}, assign terminal velocities 
  \begin{equation} \label{eqn:vrel}
  v_{\rm rel} \propto \frac{z}{\rhog}
  \end{equation}
  to each particle. 
Advect each particle vertically downward by a distance
  $v_{\rm rel}\Delta t$, where $\Delta t$ is chosen small enough that
  particles do not overtake one another. Note that the coefficient of proportionality on the right hand side of (\ref{eqn:vrel}), which depends on
quantities such as the disk mass parameter $F$ and grain properties
$s$ and $\rhos$, does not affect the shapes of the density profiles generated so long as it is the same for all particles. 

Bin particle positions and recalculate $\rhod(z)$.


	\item {\bf 1D code: Repeat (1) + (2) until the midplane
		 dust density $\rhod (0)$ rises by 30\%.}

	\item {\bf 1D $\rightarrow$ 3D code: Insert results of the 1D
          code
          	 into the 3D code and run the 3D code for ten
            orbits.}  
            
            Let $z_{\rm max}$ be the position of the highest
          particle in the 1D code, and set the dimensions of the
          shearing box in the 3D code to be $(L_r,L_\phi,L_z) =
          (1.455, 2.91, 4)\zmax$, resolved by $(N_r,N_\phi,N_z) =
          (32,64,128)$ gridpoints.\footnote{These choices imply that
            every $z_{\rm max}$ length in the $r$ and $\phi$
            directions is resolved by 22 grid points. In the $z$
            direction we need $N_z = 128$ points to achieve comparable
            resolution because the vertical grid differs from the
            horizontal grid (see section 2.4 of Paper I). 
            As discussed in \S\ref{ssec:bbr}, we test the robustness
            of our results to box size by using bigger boxes as the
            marginally stable state is approached.}  Initialize the 3D
          code by assigning $\mu(z)$, as calculated by the 1D code, to
          each horizontal gridpoint ($r,\phi$).\footnote{The transfer
            of $\mu(z)$ from the 1D code to the 3D code involved some
            smoothing because the vertical grid for the 3D code is
            $\sim$$10 \times$ coarser than that of the 1D code. We
            captured all features of the 1D dust profile $\rhod (z)$
            to within $\sim$10\% for $z \lesssim 0.8\zmax.$ Fractional
            errors generally increased away from the midplane and were
            largest at $\zmax$ where the dust content goes to zero.}
          The 3D code initializes the remaining variables---velocity,
          gas density, temperature, and enthalpy---to ensure dynamical
          equilibrium; see section 2.2 of Paper~I.

          The background radial pressure gradient is parameterized by
          the variable $\vmax$:
	   	\begin{equation}
		-\frac{1}{\rhog}\frac{\p P}{\p r} = 2\OmegaK \vmax\, .
		\end{equation}
                We fix $\vmax = 0.025\cs$ for all
                simulations. Physically, $\vmax \sim c^2_{\rm
                  s}/(\OmegaK r)$ represents the difference in
                azimuthal velocity between pressure-supported
                dust-free gas and a strictly Keplerian
                flow.\footnote{This value of $\vmax$ coincides with
                  the standard value from Paper I, although
                  technically the minimum-mass disk model in Appendix
                  \ref{app:background} gives $\vmax/\cs = 0.036$.
                  Both values of $\vmax$ correspond to cool disks
                  passively heated by their central stars (e.g.,
                  \citealt{chianggoldreich97}). Smaller $\vmax$
                  results in thinner and denser dust layers and thus
                  promotes the formation of planetesimals.
                  See \S\ref{ssec:superlinear} and Appendix
                  \ref{app:superlinear} for how our results depend on
                  $\vmax$.}

           Before running the 3D code, perturb $\mu(r,\phi,z)$ by an amount  
	   	\begin{eqnarray}
		 \Delta \mu (r,\phi,z) &=& A(r,\phi) \mu(z) [ \cos(\pi z / 2 \zmax)\nonumber \\ 
		&& +  \sin(\pi z/2 \zmax) ] ,
		\end{eqnarray}
	   where $A(r,\phi)$ is a random variable constructed in
           Fourier space (see the discussion following equation 31 of
           Paper~I). Fix the root-mean-square of the perturbations to
           be $A_{\rm rms} \equiv
           \langle A^2 \rangle ^{1/2} = 10^{-3}$.

           Run the 3D code for ten orbits.

	   \item {\bf 3D code: Assess stability.  Extend simulations
	   	 beyond ten orbits as necessary to make this assessment.}
              
              Label the dust profile ``KH-unstable" if the horizontally averaged dust-to-gas ratio at the midplane
		$$\label{eqn:mucrit} \langle \mu (z=0) \rangle\ \text{as a function of $t$}
		$$
		decreases by more than 15\%. Otherwise,
 monitor the horizontally averaged vertical kinetic energy at the midplane:
		$$\label{eqn:kecrit} \langle \mu v^2_z (z=0)\rangle/2\ \text{as a function of $t$}.
		$$
		If $\langle \mu v^2_z \rangle /2$ monotonically decreases or levels off, label the dust profile ``KH-stable.'' If $\langle \mu v^2_z \rangle/2$ is increasing towards the end of the simulation, extend the integration an additional ten orbits and re-assess stability. Repeat step (5) as necessary.
		
\item {\bf If ``KH-unstable,'' stop.}
	
Identify the last KH-stable dust
  profile, generated in the iteration just previous to that of the KH-unstable simulation, as the ``marginally
  stable state.''\footnote{This marginally stable state will be superseded by the marginally stable state identified under an improved scheme
in \S\ref{ssec:weight}.}

\item {\bf If ``KH-stable,'' pass results of the 3D code
 back to the 1D code and return to step (1).}

Fit a polynomial $\mu_{\rm poly}(z)$ to the final, horizontally averaged dust-to-gas
ratio $\langle \mu(z) \rangle$ as calculated by the 3D code. Adjust
the order of the polynomial to capture all features of the profile. If
one polynomial is insufficient, use two to create a piecewise
function. Convert $\mu_{\rm poly}(z)$ to $\rhod(z)$ by assuming the
gas profile to be Gaussian (see footnote \ref{foot:muchado}):
$\rhod(z) = \mu_{\rm poly}(z)\cdot\rho_{\rm g0} \exp[-z^2/(2 H^2_{\rm g})]$. 
Using this $\rhod(z)$, return to step (1) for the next iteration.

\end{enumerate}

Our method and results apply to any location in a disk of any mass
(arbitrary $r$ and $F$), provided our input assumptions that
self-gravity is negligible and $\vmax/\cs = 0.025$ are satisfied. They
also apply to any particle size to the extent that the disk's solid
mass is concentrated in particles of a single size (so that $z$
and $\rhog$ uniquely determine $v_{\rm rel}$; equation \ref{eqn:vrel}),
and to the extent that such particles are undisturbed by streaming
instabilities (\S\ref{sec:introduction}; \S\ref{ssec:future}). Another
way of saying all this is to note that our calculations are carried
out in dimensionless units.


\section{RESULTS}
\label{sec:results}

To orient the reader, in Figure \ref{fig:1Dcode} we show results
obtained from the 1D code only.  The dust is initially well mixed with
gas at solar metallicity ($\mu_{\rm init} = 0.015$).  As dust settles
and the
midplane dust-to-gas ratio $\mu_0$ increases, sharp cusps appear at the
edges of the dust layer where particles pile up vertically.
Pileups occur because particle fluxes $\rhod |v_{\rm rel}| \propto \mu |z|$
increase with increasing height $|z|$.
This follows from our assumption that $\mu_{\rm init}$ is
constant, which as noted at the beginning of \S\ref{sec:method}
may not be realistic.

Unlike us, \citet{garaudlin04} did not find vertical pileups at the
edges of their layer because
they chose their initial dust profile to have a scale height
equal to $0.1 H_{\rm g}$. Their initial $\mu$ profile
decreased with $|z|$ more quickly than $1/|z|$, and thus
did not satisfy the condition for pileups.
We verified this by inserting their initial profile into our 1D code.

The shapes of the settled dust profiles $\mu(z)$ and their relative
spacing in time are independent of the dust internal density
$\rho_{\rm s}$, dust particle size $s$, and the scaling parameter $F$
for disk mass.  Changing these parameters only alters the absolute
physical time elapsed (equation \ref{eq_tz}). Relative time is tracked
by the dimensionless parameter $f \equiv t/t_{\rm settle}$, labeled on
this and many subsequent figures.

Below we compare these 1D-only results to those that include
the full 3D dynamics. The solar metallicity case is described
in \S\ref{ssec:solar}. The metal-rich case ($\mu_{\rm init} = 0.06$)
is presented in \S\ref{ssec:mr}.

\subsection{Solar Metallicity}
\label{ssec:solar}

Figure \ref{fig:solarmoney} traces the evolution of dust that starts
well mixed with gas at solar metallicity. Plotted are several
KH-stable curves from the 3D code resulting from step (5) of our
procedure. For ease of comparison with the purely 1D results, the
relative timestamps in Figure \ref{fig:solarmoney}, measured by $f$,
coincide with those in Figure \ref{fig:1Dcode}. The leftmost curve at
$f = 1.0$ represents the marginally stable state
identified using our standard procedure. This state achieves a
midplane dust-to-gas ratio of $\mu_0 = 2.45$, about an order of
magnitude below the value required for gravitational instability
(equation \ref{eqn:muToomre}). In \S\ref{sec:ext} we extend our
procedure to see if we might achieve still higher dust-to-gas ratios.

Comparing Figures \ref{fig:1Dcode} and \ref{fig:solarmoney}, we see
that the (possibly unrealistic) pileups at the edges of the dust layer do not 
survive in the dynamical 3D code. By $f \approx 0.44$, the pileups are
nearly gone. At this point, the vertical extent of the dust
layer $z_{\rm max}$ has shrunk to $\sim$$0.1 \Hg$, and 
the only pileup present is the one at the midplane.

The instability that eliminates the pileups at the edges of the layer
is likely related to the Rayleigh-Taylor instability (RTI), triggered
by heavy fluid lying on top of lighter fluid, and we will refer to it
henceforth as such. The RTI originates locally at the edges of the
dust layer.  By contrast, the midplane is relatively stable (at least
until the marginally stable state is reached).  Another way of seeing
this is to note that midplane dust-to-gas ratios $\mu_0$ in Figures
\ref{fig:1Dcode} and \ref{fig:solarmoney} agree to within 25\%. Closer
examination reveals that those in Figure \ref{fig:solarmoney}
are consistently higher. This suggests that the RTI transfers some of
the dust in the pileups to the midplane.

Figure \ref{fig:solar67} confirms this transfer mechanism. The top
middle panel shows that over the course of a 20-orbit-long 3D
simulation (iteration \#6, occurring at a time $f=0.31$, out of a
total of 19 iterations), dust is redistributed from the layer's edges
to the midplane, raising $\mu_0$ by about 20\%.  Note that the effect
of the RTI has been to transport dust toward the midplane, not to
higher altitudes. The RTI is confined to where dust is unstably
stratified (increasing total density in the direction opposite to
gravity).

Compare this behavior with that in the top row of Figure
\ref{fig:solar1617}, which documents a later iteration, \#16. The top
middle panel shows that an instability has occurred near the edges of
the dust layer. Dust is redistributed to higher, not lower,
altitudes. The midplane is not affected. The instability at this
relatively late stage of settling is probably driven by the vertical
shear associated with strong density gradients at the edges of the
layer, and we will refer to it henceforth as the Kelvin-Helmholtz
instability (KHI). As a result of the KHI, gradients in density and
velocity are reduced.

The marginally stable state identified using our standard procedure
 is displayed in Figure \ref{fig:solar19101930}.
The bottom panels show that during the last iteration \#19,
the usual 30\% increase in the midplane
$\mu_0$ (left bottom) results
in a KH-unstable profile (middle bottom). In the top
panels, we redo iteration \#19, this time incrementing $\mu_0$
by only 10\% (left top). The resultant profile
is KH stable (middle and right top panels), and has 
$\langle \mu_0 \rangle = 2.45$. In \S\ref{ssec:weight},
we modify our standard procedure and extend it to later
times to achieve still higher dust-to-gas ratios in stable flows.

The $\mu$-profiles in Figures
\ref{fig:solarmoney}--\ref{fig:solar19101930} betray oscillations just
inside the edges of the dust layer.  We believe these ripples are
artificial because when each first appears, it spans only a few grid
points of the 3D code: see the $f = 0.054$ profile of Figure
\ref{fig:solarmoney}, which shows two nascent ripples. The features
probably arise because the truncated Chebyshev series used to model
the flow in $z$ has too few terms to adequately capture the steep
vertical density gradient \citep{gibbs1898}. Originating in the 3D
code, the ripples are then amplified as mini-pileups in the 1D
code. We could have tried to smooth away these oscillations by
reducing the order of our polynomial fit (step 7 of our procedure),
but chose instead to retain all features of the dust profile generated
by both codes to minimize bias.  In any case the oscillations are
eventually erased by instabilities during the later stages of settling
(Figure \ref{fig:solarmoney}). In and of themselves the oscillations
do not appear to introduce instabilities, which as discussed above are
triggered instead by smooth density gradients---realistically
computed---at the boundaries of the layer (top rows of Figures
\ref{fig:solar67} and \ref{fig:solar1617}).

\subsection{Metal-Rich Case: 4 $\times$ Solar Metallicity}
\label{ssec:mr}

\begin{figure}
\plotone{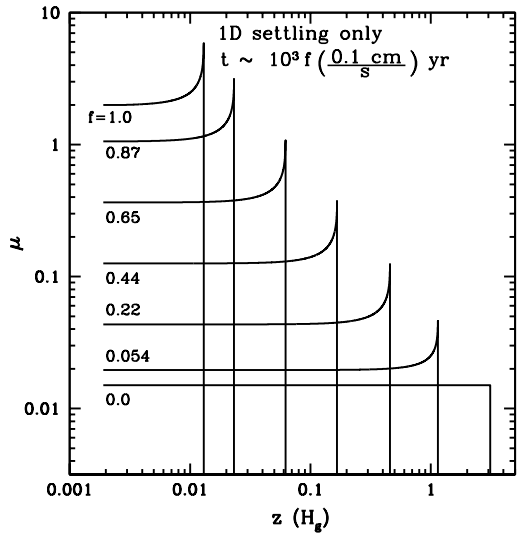}
\caption{Snapshots of settling dust computed with the 1D code
  only. Plotted is the dust-to-gas ratio as a function of height at
  various instants of time. Relative timestamps are assigned by the
  non-dimensional parameter $f$; see the inset equation for the
  absolute elapsed time, which assumes $\rhos = 1$ g/cm$^3$ and $F =
  1$ (equation \ref{eq_tz}).  Note that the shapes of these profiles
  and their relative spacing in time are independent of the absolute
  elapsed time, and thus independent of $\rhos$, $F$, and $s$. At
  $f=0$, dust begins well mixed at solar metallicity
  \citep[$\mu=0.015;$][]{lodders03}. By $f = 0.054$, a pileup has
  formed at the dust layer's edge.  The pileup is an artifact
  of our assumption that the initial dust-to-gas ratio
  $\muinit$ is constant everywhere, which causes
  particle fluxes $\rhod |v_{\rm rel}| \propto \mu |z|$ to
  decrease with decreasing height $|z|$. \\
}
\label{fig:1Dcode}
\end{figure}

\begin{figure}
\plotone{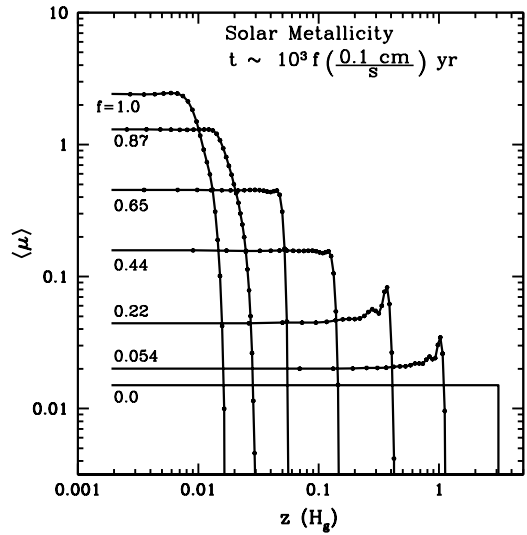}
\caption{Snapshots of settling dust computed using the standard procedure
  of \S\ref{sec:method} which combines the 1D and 3D codes, for the
  case of bulk solar metallicity.  Elapsed
  time is marked by $f$; plotted values coincide with those in Figure
  \ref{fig:1Dcode}. The shapes of the profiles and their relative
  spacing in time do not depend on the absolute elapsed time; they are
  independent of $\rhos$, $F$, and $s$.  Dust begins well mixed with gas at
  $\mu = 0.015$ and ends in the 
  marginally stable state with midplane $\mu_0 = 2.45$. Vertical
  gridpoints from the 3D code are plotted as dots. In comparison to
  the purely 1D results of Figure \ref{fig:1Dcode}, the pileup at the
  layer's edge is smoothed away, probably by the Rayleigh-Taylor instability, 
  between $f = 0.22$ and $f = 0.65$. Except for transferring some dust at altitude to the
  midplane, the instability leaves the midplane relatively
  unaffected, which until $f = 1.0$ evolves much as it does in Figure
  \ref{fig:1Dcode}. \\} 
\label{fig:solarmoney}
\end{figure}

Figure \ref{fig:mrmoney} follows the evolution of dust that is initially
well mixed with gas at $4\times$ solar metallicity. It shares
the same timeline as Figures \ref{fig:1Dcode} and
\ref{fig:solarmoney}. Thus the last profile marked $f=1.1$ in Figure
\ref{fig:mrmoney} is attained at a time 10\% later than that marked
$f=1.0$ in the other figures. This last profile is the 
marginally stable state identified using our standard procedure, for
the case of supersolar metallicity. It achieves
 a midplane dust-to-gas ratio
of $\mu_0=20.3$---large enough to exceed the Toomre threshold
in a disk that has twice the gas content of the
minimum-mass solar nebula ($F=2$ in equation \ref{eqn:muToomre}).

\begin{figure}
\plotone{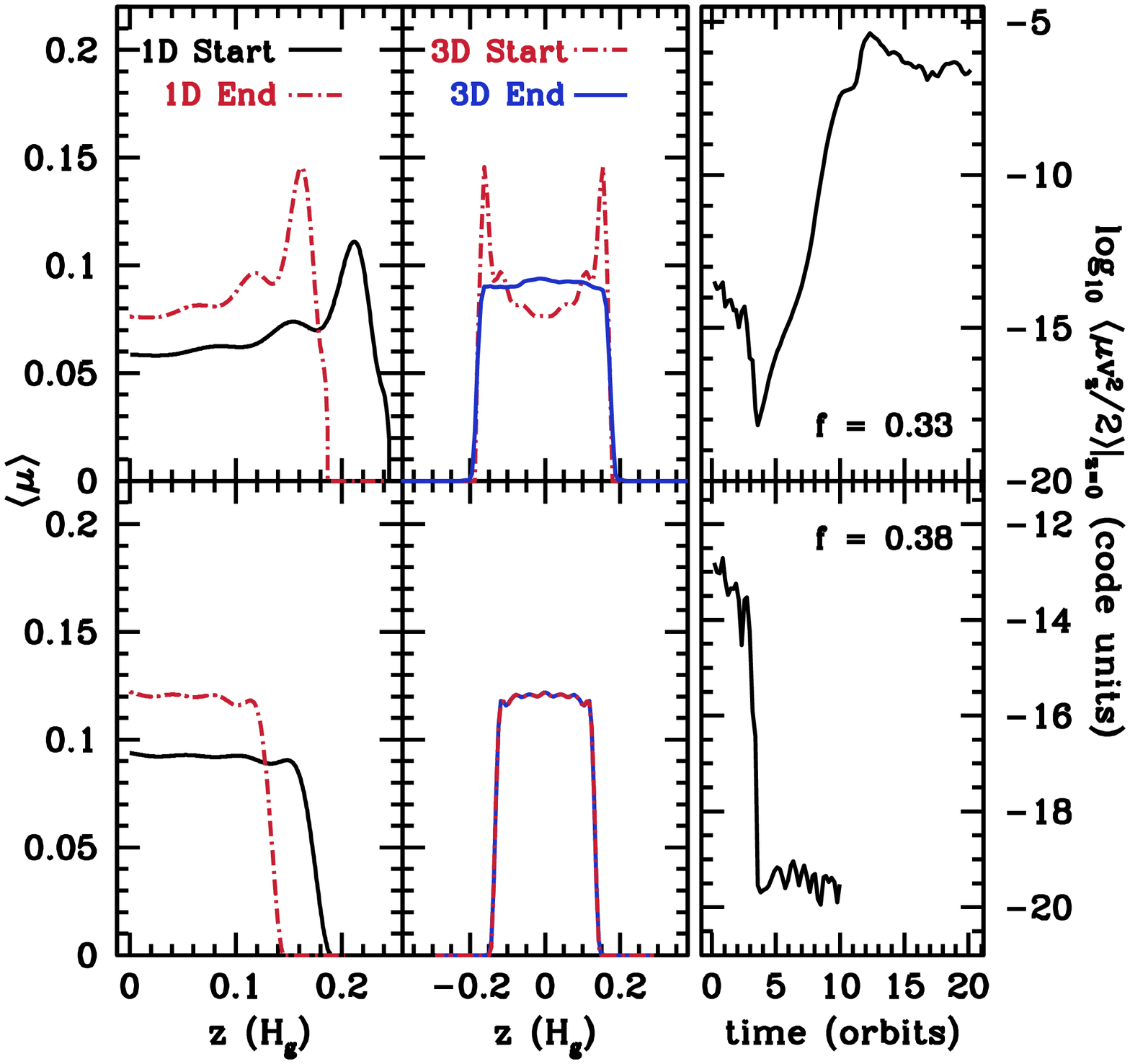}
\caption{
Two successive iterations of our procedure of
  \S\ref{sec:method}, for the case of bulk solar metallicity.
  From left to right, the panels show a starting dust profile
  (black curve) settled by the 1D code until its midplane $\mu_0$
  increases by 30\% (red dot-dashed curve). This settled curve
  is then passed to the
  3D code and evolved (blue curve) until it stabilizes (rightmost panel
  showing how the vertical kinetic energy at the midplane eventually
  levels off). Top panels show iteration \#6 of 19 (equivalently $f=0.33$ on the
timeline of Figure \ref{fig:solarmoney}). The unstably stratified
pileups collapse around $t\sim 11$ orbits, increasing
the midplane dust content by $\sim$20\% (top middle). 
Bottom panels show iteration \#7 ($f = 0.38$) which begins where
iteration \#6 leaves off---except that the kinetic energy of the flow
is reset to a low value (bottom right versus top right panels), and
the slight asymmetry in $\langle \mu \rangle$ about $z=0$ (top middle
panel, blue curve) is dropped upon fitting a polynomial only to $z \ge
0$ (bottom left, black curve). The oscillations in the $\mu$-profiles
are artifacts of having too few basis functions in $z$. They did not
seem to introduce instability, which always occurred instead at the
edges of the dust layer where gradients were steepest and
realistically computed.
}
\label{fig:solar67}
\end{figure}

\begin{figure}
\plotone{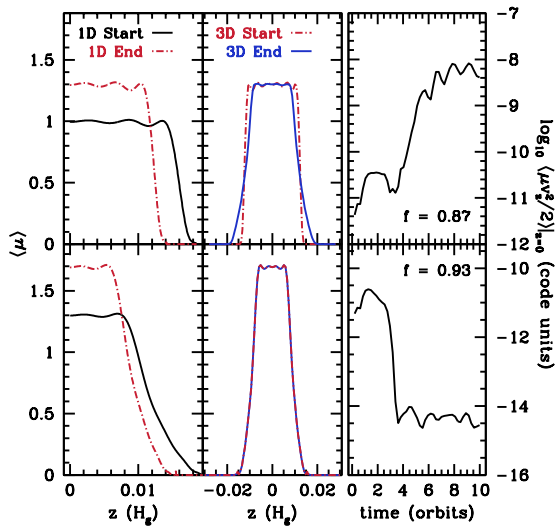}
\caption{Similar to Figure \ref{fig:solar67} but showing iterations
  \#16 (top) and \#17 (bottom) out of a total of 19, for the case of
  bulk solar metallicity. In iteration \#16, dusty gas at the layer's edges
  mixes with dust-poor gas at higher altitudes (top middle), probably
  by the KHI. The subsequent evolution
  during iteration \#17 shows no sign of instability after 10 orbits.}
\label{fig:solar1617}
\end{figure}

\begin{figure}
\plotone{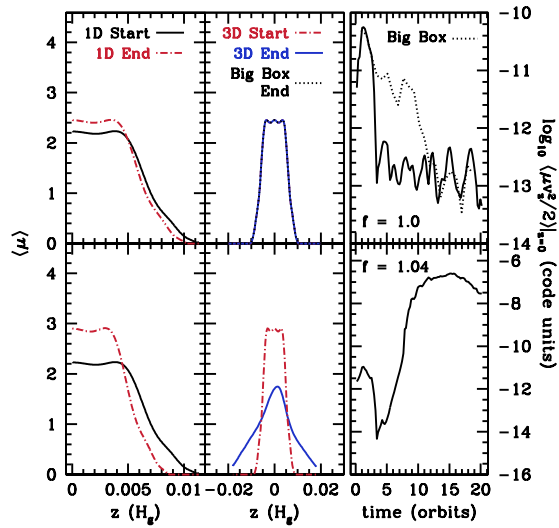}
\caption{Similar to Figures \ref{fig:solar67} and \ref{fig:solar1617}
  but showing the last couple iterations (\#19a and \#19b) which
  provisionally identify the marginally stable state for the case of
  bulk solar metallicity. Increasing the midplane dust content from
  iteration \#18 by 30\% (bottom panels) leads to a KH-unstable profile, while an increase of 10\%
  preserves KH stability (top panels). Quadrupling $L_\phi$ and
  $N_\phi$ simultaneously (dotted lines) does not change our
  answer. The marginally stable state in the top panels is refined
  according to a modified procedure in \S\ref{ssec:weight}.}
\label{fig:solar19101930}
\end{figure}

\begin{figure}
\plotone{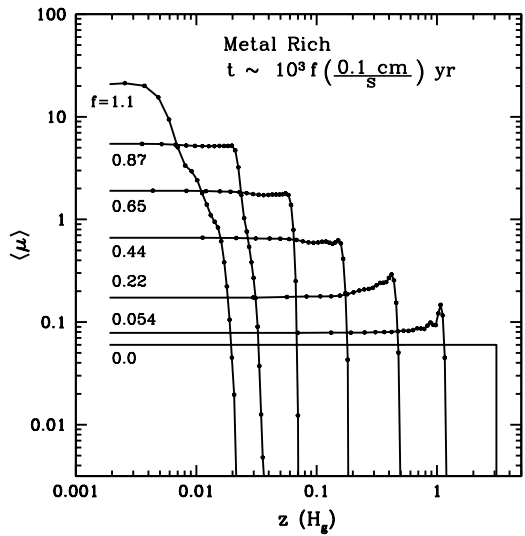}
\caption{Snapshots of settling dust computed with the full procedure
  of \S\ref{sec:method} which combines the 1D and 3D codes, for the
  case of 4$\times$ bulk solar metallicity.  Elapsed time is marked by
  $f$, measured on the same timeline characterizing Figures
  \ref{fig:1Dcode} and \ref{fig:solarmoney}.  The shapes of the
  profiles and their relative spacing in time do not depend on the
  absolute elapsed time; in this sense the evolution is not sensitive
  to $\rhos$, $F$, and $s$.  Vertical gridpoints from the 3D code are
  plotted as dots.  Dust begins well mixed with gas at $\mu = 0.06$
  and ends in the marginally stable state with midplane $\mu_0 = 20.3$.
  The midplane density in this last state already exceeds the
  threshold for Toomre instability in a disk with twice the gas
  content of the minimum-mass solar nebula (equation
  \ref{eqn:muToomre} with $F=2$).}
\label{fig:mrmoney}
\end{figure}

\begin{figure}
\plotone{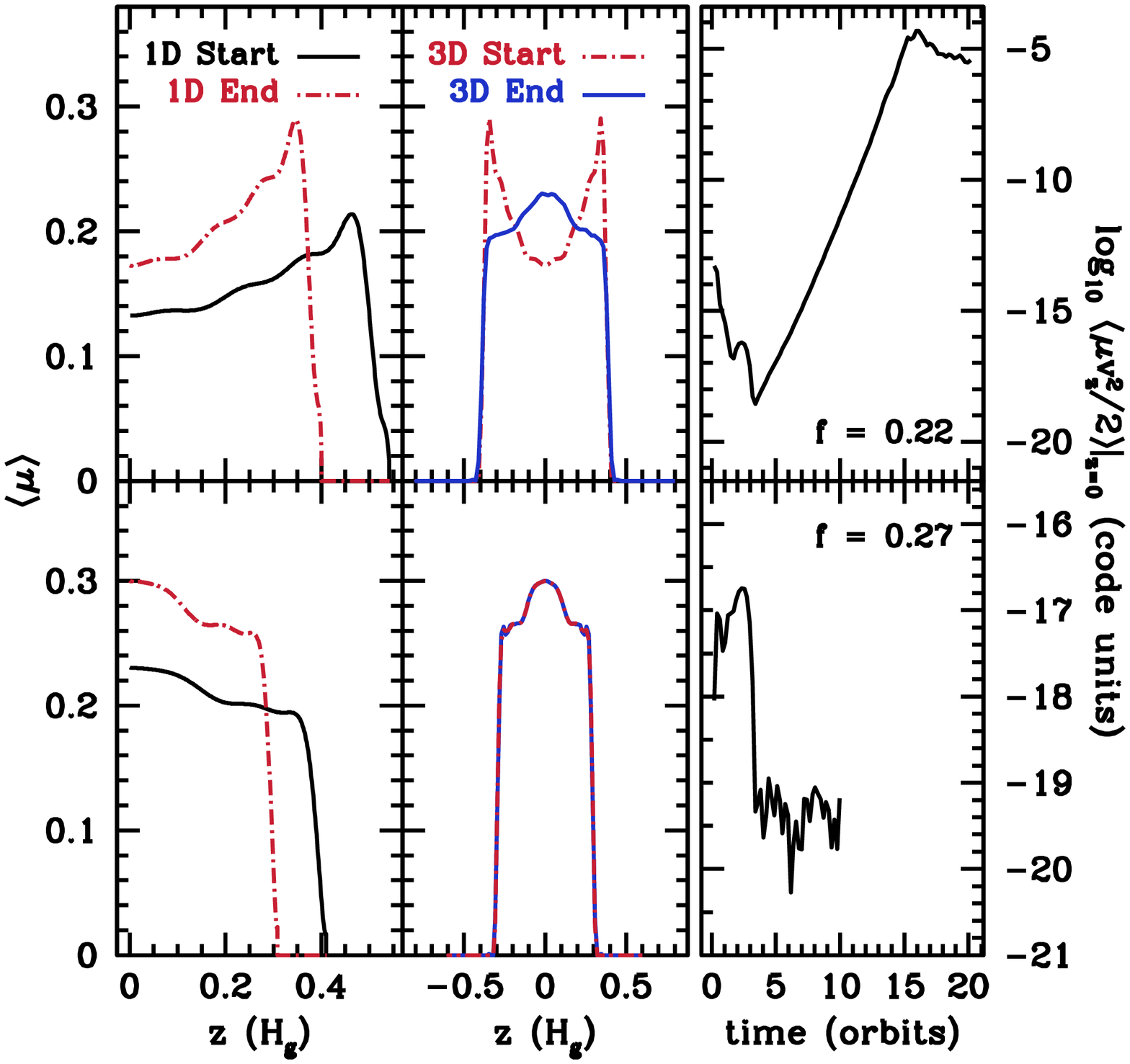}
\caption{Two successive iterations of our procedure outlined in
  \S\ref{sec:method}, for the case of $4\times$ solar metallicity.
  From left to right, the panels show a starting dust profile (black
  curve) settled by the 1D code until its midplane $\mu_0$ increases by
  30\% (red dot-dashed curve). This settled profile is then passed to
  the 3D code and evolved (blue curve) until it stabilizes (rightmost
  panel showing how the vertical kinetic energy at the midplane
  eventually levels off). Top panels show iteration \#4 of 21
  (equivalently $f=0.22$ on the timeline of Figure
  \ref{fig:mrmoney}). When the unstably stratified pileups collapse,
  they increase the dust content of the midplane by $\sim$35\% (top
  middle panel). The resultant dust profile, settled further in
  iteration \#5 (bottom panels), remains free of instabilities after
  10 orbits.}
\label{fig:mr45}
\end{figure}

\begin{figure}
\plotone{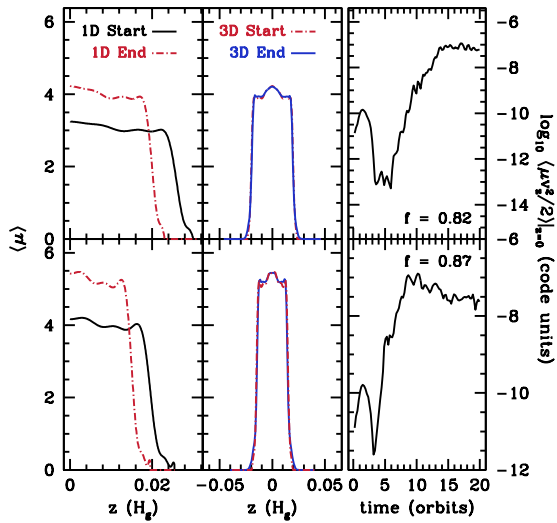}
\caption{Similar to Figure \ref{fig:mr45} but showing iterations \#15
  and \#16 out of a total of 21 for the case of $4\times$ solar
  metallicity.  Shown are two examples of KH-stable profiles whose
  midplane vertical kinetic energies
  end orders of magnitude above their starting values. Every 3D
  simulation starting with iteration \#13 in the metal-rich case shows
  this kind of sustained motion even though the density profiles may be
  KH stable according to our criterion.}
\label{fig:mr1516}
\end{figure}

\begin{figure}
\plotone{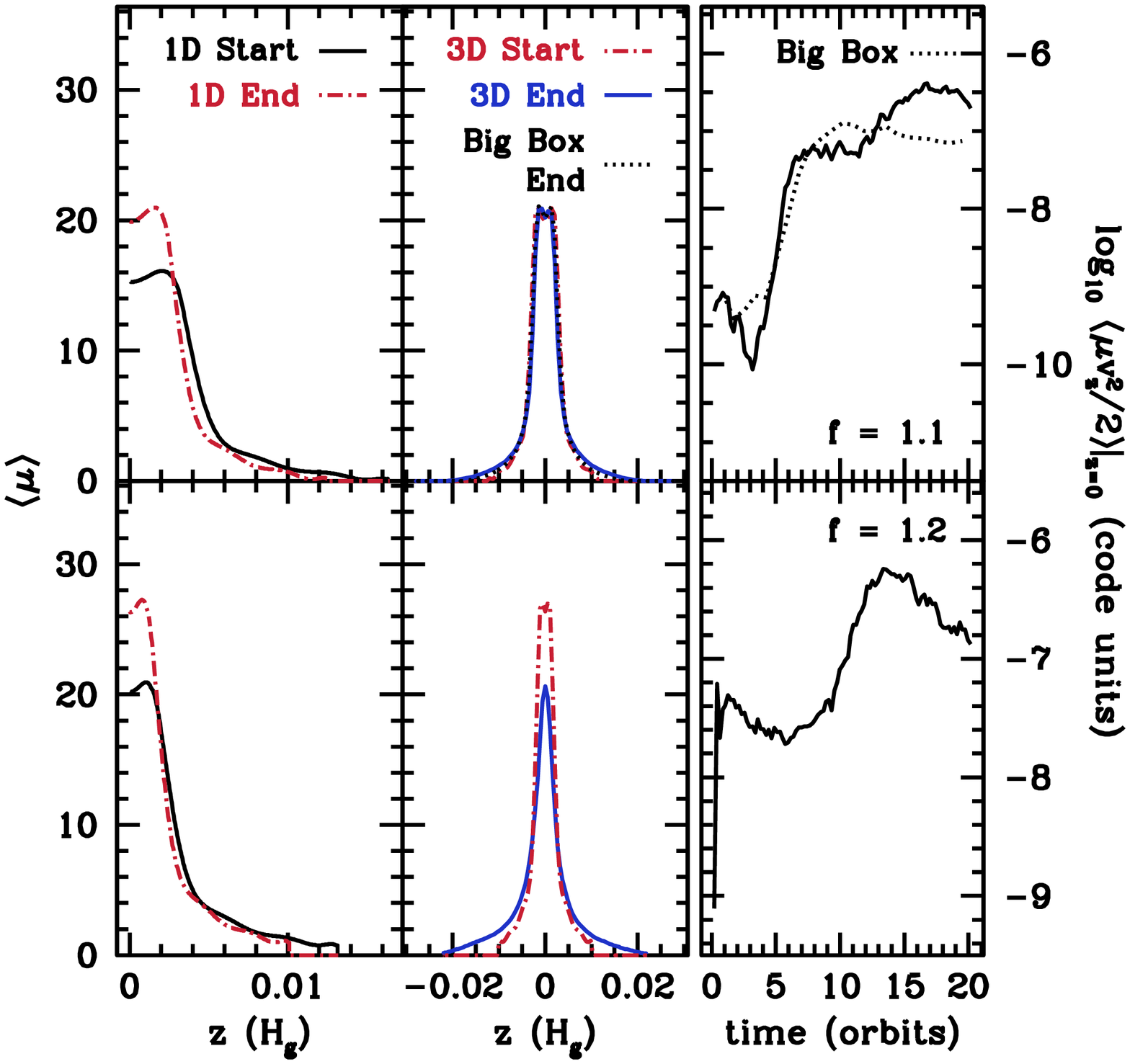}
\caption{Similar to Figures \ref{fig:mr45} and \ref{fig:mr1516} but
  showing iteration \#21 in the top panels, in which the marginally
  stable state is found for the case of $4\times$ solar metallicity
  according to our standard procedure. The midplane $\langle \mu_0
  \rangle = 20.3$, corresponding to a midplane density that exceeds
  that required for gravitational instability in a disk having twice
  the gas content of the minimum-mass solar nebula.  The same dust
  profile inserted into a shearing box four times as wide in the
  azimuthal direction as our standard box and having four times
  as many azimuthal grid points yields qualitatively the
  same result (dotted line). Settling still further according to our
  standard procedure results in KH instability (bottom panels), but in
  \S\ref{ssec:weight} we experiment with a modified procedure that
  tries to hold off KH instability for a while longer.}
\label{fig:mr2122}
\end{figure}

The evolution of the metal-rich disk over the course of a total of 21
iterations---some of which are sampled in Figures
\ref{fig:mr45}--\ref{fig:mr2122}---is similar to that of the solar
metallicity disk, with two notable differences. When the unstably
stratified pileups of dust collapse (iteration \#4, shown in the
top row of Figure \ref{fig:mr45}), enough dust is transferred to the midplane
that $\mu$ attains an appreciable maximum there.
This bump contrasts with the nearly flat profile seen
for the solar metallicity run (Figure \ref{fig:solar67}), and persists 
at least through iteration \#16 (Figure
\ref{fig:mr1516}).  A second difference is that in every KH-stable
simulation following iteration \#13, the vertical kinetic energy,
although it eventually levels off, ends orders of magnitude higher
than where it began (Figure \ref{fig:mr1516}, and top row of Figure
\ref{fig:mr2122}).  Some currents and/or turbulence appear to be
sustained as the state of marginal stability is approached. 

Related to this second point, we should acknowledge that our standard
procedure ignores whatever velocities are present at the end of a
given 3D simulation when initializing the velocities of the subsequent
3D simulation. That is, with every iteration, velocities are set anew
according to equation (\ref{eqn:vphi}), with vertical and radial
velocities reset to zero. The assumption we make
in our standard procedure is that whatever velocities are maintained
in a KH-stable layer do not stop dust from settling at the local
terminal velocity $v_{\rm rel}$.  A crude attempt at relaxing this
assumption is made in \S\ref{ssec:weight}.


\vspace{0.2in}
\section{EXTENSIONS}
\label{sec:ext}

In \S\ref{ssec:bbr} we test the robustness of our results against the
size of our simulation box. In \S\ref{ssec:weight} we modify the
procedure of \S\ref{sec:method}, pushing to still higher dust-to-gas
ratios at the midplane and revising our identification of marginally
stable states.

\subsection{Bigger Box Runs}
\label{ssec:bbr}

Box size can artificially affect stability because a given box can
only support modes having an integer number of azimuthal wavelengths
inside it. Thus too small a box may be missing modes that would
otherwise destabilize the layer.  To assess whether our box size is
too small, we redo the 3D simulations of our standard marginally
stable states (iteration \#19 of the solar metallicity case and
iteration \#21 of the metal-rich case), quadrupling simultaneously the
azimuthal box size $L_\phi$ and the number of grid points $N_\phi$. By
increasing both in tandem, we maintain the same resolution
$N_\phi/L_\phi$ as that of our standard runs. The results are
plotted as dotted lines in the top panels of Figures
\ref{fig:solar19101930} and \ref{fig:mr2122}. For both the solar
metallicity and metal-rich cases, the bigger box runs still yield
stable layers, just as the standard box runs do. We conclude that our
standard box sizes are probably adequate.

This conclusion is a bit surprising when we compare our standard box
size to our findings in Paper I. We found in Paper I that those KH
modes that most visibly disrupted the dust layer had azimuthal
wavelengths between $2.6\zmax$ and $4.3\zmax$. Our standard choice
here for azimuthal box size is $L_\phi = 2.91 \zmax$, which at face
value means that we are only resolving one wavelength at best of an important
mode. However, this simple comparison may not be fair.  In
Paper I we studied dust layers characterized by a spatially constant
Richardson number. The vertical density profiles there differ somewhat from
those derived here. In particular the profiles in Paper I have steep
density gradients everywhere, whereas here density gradients are steep
only at the edges of the layer. When a layer in Paper I became KH
unstable, it seemed to do so everywhere at once, whereas here
instability always originates at the edges. We have verified that this
is true even for the final iterations leading to our 
identification of the marginally stable state. Obviously these edges
have vertical thicknesses that are smaller than that of the layer as a
whole. Since the most unstable azimuthal wavelength of the KHI is expected to be of order the 
vertical thickness of the shearing layer \citep[e.g.,][]{chandra81}, it seems that our standard box sizes here, though small compared to our
box sizes in Paper I, permit us to resolve several azimuthal
wavelengths of those modes that are most able to disrupt the thin edges.

\subsection{Refining the Marginally Stable State Using a Modified Settling Procedure}
\label{ssec:weight}

Using our standard procedure of \S\ref{sec:method}, we can only
provisionally identify marginally stable dust profiles.  The
identification is provisional because by settling all dust particles
at their local terminal velocities $\vrel$, we wind up with edges so
unstable that they also destabilize the midplane. In reality, dust
particles at the edge may stop settling because they attain a state of
marginal stability first, remaining lofted up by the gas motions they
stir up locally, and leaving dust particles near the midplane free to
keep settling.  In other words, marginal stability may be reached
sequentially, starting from the edges and ending at the midplane. Our
standard procedure does not allow for this kind of gradual evolution
because the 1D code settles all dust particles at their local terminal
speeds regardless of their location.  In this sense our standard
procedure is too blunt because it does not allow for slower settling
at the stirred-up edges and faster settling at the quiescent
midplane. 
True marginally stable states should have midplane dust-to-gas ratios even higher than the maximum ones displayed in
Figures \ref{fig:solarmoney}, \ref{fig:solar19101930},
\ref{fig:mrmoney}, and \ref{fig:mr2122}.\footnote{Another reason our
  dust profiles underestimate actual dust-to-gas ratios is because we
  neglect vertical self-gravity, which enhances stability by
  increasing the Brunt frequency \citep{sekiya98,youdinshu02}.}

To remedy this shortcoming, we modify our procedure by applying a
weighting function $0 < W(z) \le 1$ to each dust particle's settling
velocity. 
Starting with a KH-stable state near the end of our standard sequence
of iterations, we continue to let dust particles drift to the midplane
in the 1D code, not at $\vrel(z)$ but at the weighted velocity $W (z)
\vrel(z)$. We use either a Fermi function
	\begin{equation}\label{eqn:winfermi} 
	W (z) = \frac{1}{1 + \exp[(z-\zhalf )/\zw ]},
	\end{equation}
described by two parameters $\zhalf$ and $\zw$, or a Gaussian 
	\begin{equation}\label{eqn:wingauss}
	W (z) = \exp(-z^2 /2z^2_{\rm w}),
	\end{equation} 
        described by a single parameter $\zw$.  The choice of
        weighting function is somewhat arbitrary; it depends on the
        shape of the dust profile to be settled and is made
        case-by-case according to considerations outlined below.
The intent of the weighting function is to slow the steepening
of density gradients at the dust layer's edges and thereby
prevent the edges from destabilizing the entire layer.

\begin{figure}
\plotone{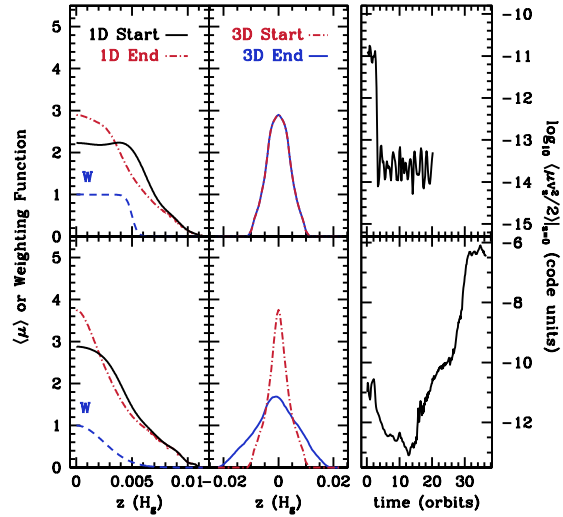}
\caption{Extended settling simulations using the modified procedure of
  \S\ref{ssec:weight}, for the case of bulk solar metallicity.
  We start with iteration \#18 from our standard procedure (black curve,
  top left).
  A Fermi weighting function with
  $\zhalf = 0.005\Hg$ and $\zw = 0.05\zhalf$ (equation
  \ref{eqn:winfermi}, labeled '$W$' at top left) allows dust near the midplane
  to settle more
  than dust at higher altitude. The settled profile attains
  a midplane $\langle \mu_0 \rangle = 2.9$ and is KH stable (top middle
  and right panels). Further settling, this time
  using a Gaussian weighting function with $\zw = 0.0025 \Hg$,
  results in KH instability (bottom panels). Although the modified procedure enables
  us to settle beyond the last stable state identified
  using our standard procedure, the gains are not large enough
  to reach the Toomre density in solar metallicity disks.}
\label{fig:solarwin}
\end{figure}

We start with the KH-stable profile in iteration \#18 of our solar
metallicity sequence (black solid curve in top left panel of Figure
\ref{fig:solarwin}).  The dust layer is characterized by a ``core''
from $z = 0$ to $z \approx 0.5 \zmax$ over which $\langle \mu \rangle$
is fairly constant, and an ``edge'' from $z \approx 0.5 \zmax$ to $z =
\zmax$ over which the dust content drops to zero.  Because the
instabilities that threaten to disrupt the layer originate in the edge
and not the core, we seek a weighting function $W(z)$ that slows the
downward drift of dust in the edge but not in the core.  At the same
time $W(z)$ should not be so strongly weighted toward the midplane
that the core disconnects from the edge and opens a local gap in dust
content.  We find upon experimentation that a Gaussian does not have
enough flexibility to meet these requirements for this particular
iteration.  However a Fermi function with $\zhalf =
0.005\Hg$---corresponding approximately to the boundary of the
core---and $\zw = 0.05\zhalf$ seems to work well (blue dashed curve in
top left panel of Figure \ref{fig:solarwin}). We use this weighting
function to settle the dust profile until its midplane $\mu_0$ increases
by 30\% to a value of 2.9 (red dot-dashed curve).  This settled layer
remains KH stable (top middle and right panels of Figure
\ref{fig:solarwin})---unlike the layer settled without the weighting
function (bottom panels of Figure \ref{fig:solar19101930}).

In the new profile settled with our modified procedure,
the distinction between the core and the edge is no longer so sharp.
Thus to settle this new profile even further, a simple Gaussian for
the weighting function seems to suffice.  Choosing $\zw = 0.0025 \Hg
\approx 0.25 \zmax$, we attempt to increase the midplane $\mu_0$ yet
again by 30\%, but find the resultant profile to be KH unstable
(bottom panels of Figure \ref{fig:solarwin}).

\begin{figure}
\plotone{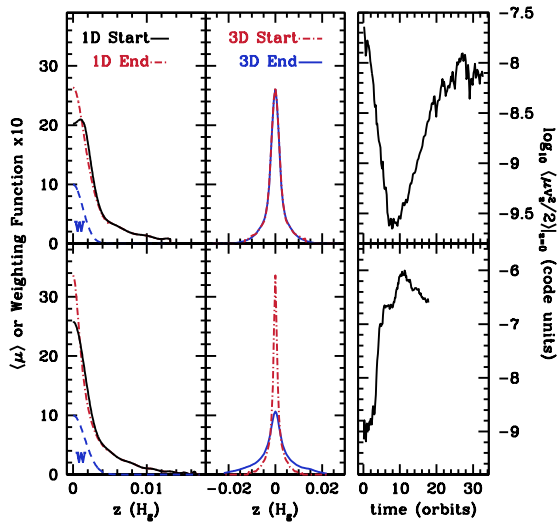}
\caption{Extended settling simulations using the modified procedure 
  of \S\ref{ssec:weight}, for the case of $4\times$ bulk solar
  metallicity.  We start with iteration \#21 from our standard
  procedure (black curve, top left). A Gaussian weighting function
  with $\zw = 0.00132\Hg = \zmax/12$ is used to settle preferentially
  the innermost layers, which achieve a maximum $\langle \mu_0 \rangle =
  26.4$ and remain KH stable (top middle and right panels).  Although
  further gains in $\mu_0$ did not materialize (bottom panels, using a
  Gaussian of $\zw = 0.00127\Hg$), $\mu_0$ is already high enough
  that gravitational instability is viable in a disk having $\sim$1--2 times
  the gas content of the minimum-mass solar nebula.
}
\label{fig:mrwin}
\end{figure}

Similar results are obtained for the metal-rich case (Figure
\ref{fig:mrwin}).  Using Gaussian weighting functions we are able to
push the midplane dust-to-gas ratio $\mu_0$ to a new record of 26.4, which is
30\% greater than the value attained using our unweighted standard
procedure.


\section{SUMMARY AND DISCUSSION}
\label{sec:dissum}

To form rocky planets and gas giant cores, dust must amass in a
circumstellar disk. In the classic scenario for forming planetesimals,
dust settles vertically toward the midplane into an ever thinner and
denser layer that eventually becomes gravitationally unstable
\citep{safronov69, goldreichward73}.  Toomre's criterion for
gravitational instability (GI) is satisfied for
midplane dust-to-gas ratios $(\rhod/\rhog)_0 \equiv \mu_0 \gtrsim
\mu_{\rm 0,Toomre}$, where $\mu_{\rm 0,Toomre} \approx 34$ for a
minimum-mass nebula orbiting a solar-mass star (equation
\ref{eqn:muToomre}; note that $\mu_{\rm 0,Toomre}$ is nearly independent
of disk radius).  For comparison, in a disk of well-mixed dust and
gas at solar abundance, $\mu_0 \approx 0.015$ \citep{lodders03}. Whether
dust can accumulate until its density increases by more than three
orders of magnitude depends on how turbulent the ambient gas is. Even
supposing that gas in certain regions of the disk is not intrinsically
turbulent (e.g., because it is too weakly ionized to support the
magnetorotational instability), the dust itself may excite turbulence
in gas by a Kelvin-Helmholtz-type shearing instability (KHI). The KHI
is triggered when the velocity gradient between dust-rich gas at the midplane
and dust-poor gas at altitude becomes too large. Barring gravitational
instability, dust should settle to a state that is marginally stable
against the KHI. The question of whether gravitational instability is
viable translates into the question of whether the state that is
marginally stable to the KHI has $\mu_0 \gtrsim \mu_{\rm 0,Toomre}$
(this is a necessary but not sufficient criterion for the
formation of planetesimals by gravitational collapse;
see footnote \ref{foot:slow}).

In this paper, we sought out such marginally stable states by
numerical simulation.  Starting with dust well mixed with gas at
either bulk solar or supersolar metallicity, we allowed dust to settle
vertically until dynamical instabilities prevented the midplane
density from increasing further. We tracked the approach to the
marginally stable state using a combination of a 1D settling code and
a 3D shearing box code, working in the limit that particles are small
enough not to excite streaming instabilities.  All the instabilities
that afflicted our dust layer originated at the layer's edges, where
dust density gradients were steepest. We found evidence for two kinds
of instabilities: the usual KHI driven by vertical shear, and the
Rayleigh-Taylor instability (RTI) driven by the weight of piled-up
dust.  These instabilities were mostly confined to the dust layer's
top and bottom surfaces, leaving dust near the midplane free to settle
but occasionally speeding up the accumulation of solids by
transferring dust from pileups downward.  The midplane density
stopped increasing when the dust layer became so thin that
instabilities at the edges threatened to overturn the entire layer.

Using our standard procedure of \S\ref{sec:method}, we attained
maximum dust-to-gas ratios of $\mu_0 \approx 2.45$ and $20.3$ for the
cases of solar and $4\times$ solar bulk metallicity, respectively
(Figures \ref{fig:solar19101930} and \ref{fig:mr2122}). These values are
lower limits because in our standard procedure dust particles at the layer's top
and bottom faces keep settling until they excite instabilities
so vigorous that dust at the midplane is stirred up.
In reality, dust at the layer's edges may reach a state of marginal
stability and stop settling, leaving dust near the midplane free to
settle further.  We modified our procedure in \S\ref{ssec:weight} to
try to account for this effect, reaching $\mu_0 \approx 2.9$ and
$26.4$ for the two metallicity cases (Figures \ref{fig:solarwin} and
\ref{fig:mrwin}).  These values are still lower limits because our
simulations omit self-gravity. But the correction for self-gravity
should be small for the solar metallicity disk, on the order of 10\%
($\sim 2.9/34$). For our supersolar metallicity disk, the correction
for self-gravity might be on the order of unity ($\sim 26.4/34$)---although
it might also be much higher, as \citet{sekiya98} and \citet{youdinshu02}
showed that vertical self-gravity can yield a singularity in $\mu_0$.

We conclude that a minimum-mass disk of bulk (height-integrated) solar
metallicity orbiting a solar-mass star cannot form planetesimals by
self-gravity alone: even neglecting turbulence intrinsic to gas, the
KHI would force the midplane dust density to fall short of the Toomre
density by about an order of magnitude.  Our results make clear what
changes to the circumstellar environment would be needed for
self-gravity to prevail.  To attain the Toomre density in a
minimum-mass gas disk, the bulk metallicity would need to be enhanced
over solar by a factor of a few $\lesssim 4$.  For disks with total
mass (gas plus dust) greater than that of the minimum-mass solar
nebula, the required degree of metal enrichment would be lower.

Our results agree with those of the prescriptive model of
\citet{weiden06}, who found that the density of mm-sized particles
($\taus \sim 0.001$) at $r = 3$ AU in a disk for which $\rhog = 1.6
\times 10^{-9} \gm \cm^{-3}$ ($F \approx 1.3$) and $\Sigmad/\Sigmag
\approx 0.015$ (solar metallicity) fell short of the Toomre density by
about a factor of 10.  When the bulk metallicity $\Sigmad/\Sigmag$
increased to $0.054$, the Toomre density was exceeded by a factor of 3.

\subsection{How Spatially Constant is the Richardson Number?}\label{ssec:howconstant}

In Paper I, as in previous works
\citep{sekiya98,youdinshu02,youdinchiang04}, all dust profiles were
assumed to have spatially constant Richardson numbers $Ri$. The dust
profiles we have computed are free of this assumption, whose validity
we can now test.

We calculate $Ri(z)$ for our marginally stable states, derived under
both standard (\S\ref{sec:method}) and modified (\S\ref{ssec:weight})
procedures. To compute the numerator (Brunt frequency) of $Ri$ in
equation (\ref{eqn:richardson}), we use the horizontally averaged
dust-to-gas ratio $\langle \mu(z) \rangle$, computing derivatives
using centered differences and assuming the gas to obey a Gaussian
density profile (see footnote \ref{foot:muchado}). To compute the
denominator (vertical shearing rate) of $Ri$, we also use $\langle
\mu(z) \rangle$, inserting it into equation (\ref{eqn:vphi}) and
computing therefrom the velocity derivative.  Of course we could also
compute the denominator more directly by using the simulation output
itself for $v_\phi$---this alternative approach turns out to give
identical results for the solar metallicity disk, but for the metal-rich
disk the $Ri(z)$ so generated varies much more erratically. As noted in
\S\ref{ssec:mr}, the metal-rich disk sustains gas motions well above
those we put in as seed noise. These motions are not strong enough to
overturn the dust layer but they are large enough to render $Ri$
highly variable, both in time and space.  By not using the simulation
data for $v_\phi(z)$ and relying instead on the better behaved
$\langle \mu(z) \rangle$, we effectively average $Ri$ in time and
space.

\begin{figure}[h]
\includegraphics[scale=0.40]{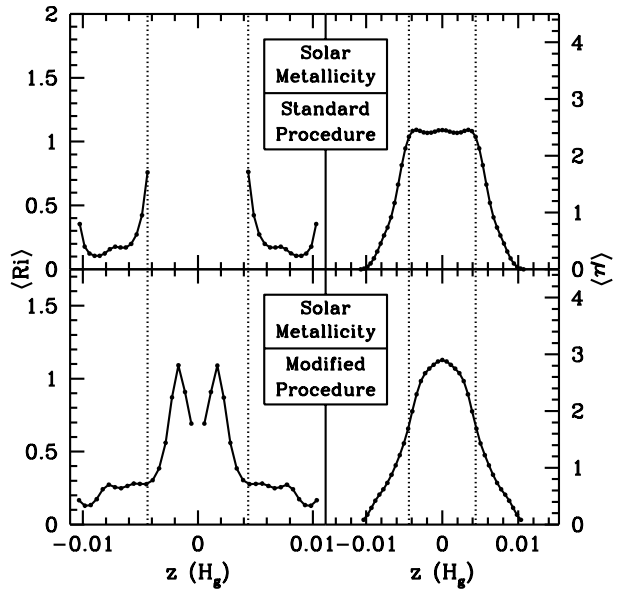}
\caption{Richardson numbers $Ri$ from the marginally stable
  profile of our standard procedure (top left) and from the 
  marginally stable profile of our modified procedure (bottom left),
  for the case of solar metallicity.
  Vertical dotted lines separate the ``core'' from the
  ``edges'' in the standard profile (top right); these dotted lines are 
  extended into the bottom panels for reference. We plot $Ri$
  everywhere except where density gradients are too small
  to compute reliably; thus we avoid the entire core region of the 
  standard profile, and the midplane of the modified profile.
  Over most of the edges---those layers outside the dotted lines
  which have directly experienced instability, almost certainly
  related to the KHI---the Richardson number varies between $\sim$0.1--0.3.
  Thus, the traditional assumption that dusty sublayers relax to states
  of spatially constant $Ri$ receives some empirical support from this figure.
}
\label{fig:solarri}
\end{figure}

Results for the solar metallicity runs are shown in Figure
\ref{fig:solarri}.  We plot $Ri$ only where $\langle \mu
\rangle$-gradients are large enough to be computed reliably---thus we
avoid regions closest to the midplane.  Although we find that $Ri$ is
not a strict constant, it varies only between 0.1 and 0.3 within a
large fraction of the edges of the dust layer---precisely where
instabilities, presumably shear-driven, have rearranged dust. This
result compares favorably with Paper I, where we found that a solar
metallicity disk has a critical Richardson number of $\Ricrit \approx
0.2$.

\begin{figure}[h]
\includegraphics[scale=0.40]{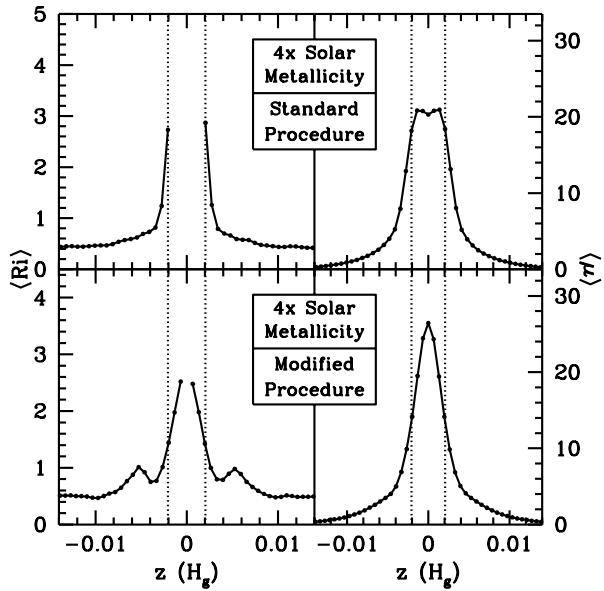}
\caption{Same as Figure \ref{fig:solarri} except for the case of $4\times$ bulk solar metallicity. Here the evidence that layers relax to states of
spatially constant $Ri$ is even stronger than for the case of
solar metallicity. Moreover, the value to which $Ri$ tends
in this metal-rich case is higher than for the solar metallicity
case: 0.4--0.6 here, versus 0.1--0.3 in Figure \ref{fig:solarri}.
This trend of increasing $Ri$ with increasing bulk metallicity $\Sigmad/\Sigmag$
is the same trend independently identified in Paper I (see Figure 6 of that
paper). In the bottom panels showing the marginally
stable profile identified using our modified procedure, the bumps
near $z\approx \pm 0.005\Hg$ are probably artificial, a reflection
of our imposed weighting function $W(z)$.
}
\label{fig:mrri}
\end{figure}

Evidence for a constant $Ri$ within the edges of the dust layer
is even stronger for the metal-rich disk, as shown in Figure
\ref{fig:mrri}. Here $Ri$ hovers near 0.5 over much of the edges---again
consistent with Paper I. See Figure 6 of that paper; admittedly
the curve for $\Ricrit (\Sigmad/\Sigmag)$ in Paper I needs to be extrapolated
to the supersolar metallicity considered here, $\Sigmad/\Sigmag = 0.06$.

The $Ri(z)$ profiles in Figures \ref{fig:solarri} and \ref{fig:mrri}
differ from those in Figure 3 of \citet{baistone10}; see the dashed
curves corresponding to their 3D simulations, all of which include marginally
aerodynamically coupled particles. These differences support their
arguments that their simulations were not afflicted by the KHI.

In summary, the assumption made in other studies that well-coupled
particles settle into a layer for which $Ri$ is spatially constant
does not appear too bad. 
This is welcome news, not least because it implies that the final
marginally stable states are robust against uncertainties
in initial conditions (\S\ref{sec:method}).
One caveat is that we have not tested this
assumption for those regions closest to the midplane, as they could
not relax by our method before being disrupted by instabilities at the
edges.
This is an area where more work can be done; see \S\ref{ssec:future}.
Another caveat, supported independently by Paper I,
is that the critical value of $Ri$ to which dust relaxes is not
unique, but increases with bulk metallicity $\Sigmad/\Sigmag$. For a
solar metallicity disk, $\Ricrit \approx 0.2$, but for one having
4$\times$ solar metallicity, $\Ricrit \approx 0.5$. This trend has not yet been explained. \\ \ \\ \

\subsection{The Super-Linear Relation Between Maximum Dust-to-Gas Ratio $\mu_0$ \\ and Bulk Metallicity $\Sigmad/\Sigmag$}\label{ssec:superlinear}

The degree to which $Ri$ is constant is related
to the scaling between the maximum midplane
dust-to-gas ratio $\mu_0$ and bulk metallicity
$\Sigmad/\Sigmag$. Naively it might be expected that $\mu_0$ scales
linearly with $\Sigmad/\Sigmag$---a greater total amount of metals
simply yields a proportionately dustier midplane---and indeed a linear
relation is implied by our order-of-magnitude estimate in equation
(\ref{eqn:muRi}). 
But we did not find a linear trend in our
simulations. We find instead that the relation is
super-linear: a factor of 4 increase in $\Sigmad/\Sigmag$ results in a
factor of 9.1 increase in maximum $\mu_0$ (26.4 versus 2.9). 

A super-linear trend is predicted by theories assuming a constant
Richardson number \citep{sekiya98,youdinshu02}. The 
large gain in midplane density afforded by a comparatively
modest increase in bulk
metallicity is partly what makes planetesimal formation by
gravitational instability so attractive. Increasing $\Sigmad/\Sigmag$
does more than just increase the total amount of metals in the
disk---it also helps to stabilize it, by decreasing the vertical shear
$\partial v_\phi/\partial z$. In the limit $\mu_0 \sim (\Sigmad/\Sigmag) \Hg / \Delta z \gg 1$, where $\Delta z$ is the dust layer thickness,
we have from equation (\ref{eqn:vphi}):
$$\frac{\partial v_\phi}{\partial z} \sim \frac{\eta \OmegaK r / \mu_0}{\Delta z} \sim \frac{\eta \OmegaK r}{\Hg} \frac{1}{\Sigmad/\Sigmag}$$
which decreases with increasing $\Sigmad/\Sigmag$. By comparison the Brunt
frequency $[(g / \rho) \partial \rho / \partial z]^{1/2} \sim
[(\OmegaK^2 \Delta z / \mu) \mu / \Delta z]^{1/2} \sim \OmegaK$ hardly
changes with $\Sigmad/\Sigmag$.  Thus the Richardson number
increases as $\Sigmad/\Sigmag$ increases; the enhanced
stability allows $\Delta z$ to decrease; and thus $\mu_0 \propto \Sigmad/\Delta z$ scales super-linearly with $\Sigmad$.

The above order-of-magnitude relations show qualitatively how a
super-linear trend follows from the decrease in vertical shear,
and the consequent increase in stability, brought about by an increase in
bulk metallicity. However, these simple relations are not enough to
quantify the super-linear trend because $\Delta z$ appears to have
cancelled out of both $\p v_{\phi}/\p z$ and the Brunt frequency. This difficulty is avoided in a more formal derivation of the relation between $\mu_0$ and $\Sigmad/\Sigmag$, made under the
assumption of constant $Ri$, as described in Appendix
\ref{app:superlinear}.

We note further that $\mu_0$ scales with the inverse of the radial pressure gradient
parameter $\vmax/\cs$ (equivalently $\eta$) in the same super-linear
way as for $\Sigmad/\Sigmag$. The smaller is $\vmax/\cs$, the greater
is the maximum $\mu_0$ attainable; the relation between these quantities
is derived under the assumption of constant $Ri$ in Appendix
\ref{app:superlinear}. Thus we expect our numerical results for $\max \mu_0$ (2.9,
26.4) to depend sensitively on our assumed value of $\vmax/\cs = 0.025$.
(\citet{baistone10b} also reported that the degree of clumping caused by
the streaming instability increased strongly with decreasing $\vmax/\cs$.)



\subsection{Future Directions}\label{ssec:future}

With each iteration of our standard procedure we allowed dust
particles to settle at their full terminal velocities, regardless of
gas motions evinced in previous iterations. We tried to account for
these gas motions in a modified procedure by reducing settling
velocities at altitude where dust may have already attained marginal
stability. Settling velocities were reduced by weighting functions
chosen by eye. This modified procedure enabled us to extend the
settling sequence by one iteration but no more. Other weighting
functions might allow the sequence to be extended 
further. Introducing weighting functions earlier in the sequence
(rather than at the end of our standard procedure, as we have done),
and increasing the midplane density by a smaller increment with each
iteration (less than the 30\% increment that we have adopted), would
allow for a more gradual evolution and possibly permit the midplane to
reach still greater densities.

Such a program would be straightforward to pursue but would be subject
to arbitrariness in the forms of the weighting functions.  A more
direct approach would be to abandon our hybrid 1D+3D scheme and
upgrade the 3D code to allow for a non-zero aerodynamic stopping time
$t_{\rm stop}$ for dust. Then both settling and stability could be
tracked within a single 3D code. Similar codes have been
written \citep[e.g.,][]{johansenetal09,baistone10a}, but their
application has been focussed on the streaming instability, on
particles having $\OmegaK t_{\rm stop} \gtrsim 0.1$ and
(model-dependent) sizes upwards of decimeters. By contrast, we are
interested in the possibility that even the smallest particles, for
which $0 < \OmegaK t_{\rm stop} \ll 1$, undergo gravitational
instability. 
The problem of settling small particles may be coupled to the problem
of settling big ones. Even if marginally coupled particles comprise
only a minority of the disk's solid mass, the turbulence they induce
by the streaming instability may prevent smaller particles from
settling into the thin layers required for gravitational instability
\citep{baistone10}. Quantifying what is meant by ``minority'' remains
a forefront issue. An efficient scheme for numerically simulating this
problem would combine the anelastic methods we have adopted (so that
the code timestep is not limited by the sound-crossing time) with an
implicit particle integrator like the kind devised by
\citet{baistone10a} (so that the code timestep is not limited by
$t_{\rm stop}$).


Adding self-gravity would be another improvement. As noted at the
beginning of \S\ref{sec:dissum}, vertical self-gravity is expected to
increase the maximum dust-to-gas ratio by of order 10\% for the case
of bulk solar metallicity. For the case of a few $\times$ supersolar
metallicity, the magnitude of the correction is
uncertain. It is probably at least of order unity, but might be much
more, given the appearance of an infinite density cusp in those
solutions of \citet{sekiya98} and \citet{youdinshu02} that account for
vertical self-gravity.  At the same time, self-gravity might actually
limit maximum dust-to-gas ratios if the fluid becomes
gravito-turbulent without producing collapsed objects
\citep{gammie01}.

\subsection{Connection to Observations and The Need For Supersolar
  Metallicity}\label{ssec:connect}

Observations have unveiled several trends between stellar properties
and the likelihood of planet occurrence. Among the most well-known is
the positive correlation between the occurrence rate of giant planets
and the host star metallicity [Fe/H] \citep{gonzalez97, santosetal04,
  fischervalenti05}. \citet{johnsonetal10} provided a comprehensive
analysis, using a sample of 1266 local stars to ask whether the trend
with metallicity persists across all stellar masses $M_\ast$. The
answer is contained in their Figure 2.  The need for supersolar
metallicity is clear for M dwarfs ($0.2 < M_\ast/M_\sun < 0.7$), where
the average metallicity of the planet-hosting stars is [Fe/H] =
0.4. Metal-rich stars presumably once carried metal-rich disks, and so
the planet-metallicity correlation for M dwarfs supports our results,
and those of others (\citealt{sekiya98}; \citealt{youdinshu02};
\citealt{leeetal10}; see also \citealt{johansenetal09};
\citealt{baistone10}) that planetesimals form much more readily in
metal-rich environments. In particular the data for M dwarfs indicate
that a mere factor of $10^{0.4} = 2.5$ increase in metallicity above
solar substantially increases the probability of planet
occurrence. This is consistent with our finding of a super-linear
trend between maximum dust-to-gas ratio and bulk metallicity
(\S\ref{ssec:superlinear} and Appendix \ref{app:superlinear}).

However, the planet-metallicity correlation weakens systematically
 with increasing stellar mass \citep{johnsonetal10}. For
A stars ($1.4 < M_\ast/M_\sun < 2.0$), the correlation is arguably
not present. This calls into question the need for supersolar metallicities
to form planetesimals.  The observations of \citet{johnsonetal10} 
might still be reconciled with gravitational instability if more
massive stars host more massive disks, although disk mass would have
to scale with stellar mass in a faster than linear way to lower the
threshold Toomre density (equation \ref{eqn:muToomre}).  The
possibility also remains that the observations are not actually a
direct or sensitive probe of the theory. The observations concern
stellar metallicity, which might at best correlate with the
global metallicity of the disk, integrated over both disk height and
disk radius.  By comparison, theory concerns the local
metallicity $\Sigmad/\Sigmag$, integrated over height but not radius.
This local metallicity (not to be confused with the local dust-to-gas ratio $\mu$) can evolve substantially from its global value,
as a consequence of radial particle drifts and photoevaporation (e.g., CY10).

Rather than look to their parent stars for evidence for local disk
enrichment, we can look to the planets themselves. If planetesimals
can only form in metal-enriched environments, we expect that the
resultant planets will also be metal-enriched. \citet{guillotetal06}
computed the bulk metallicities of the first nine extrasolar planets
discovered to be transiting, all of which are hot Jupiters.  The
results are listed in Table \ref{tab:metal}, together with the modeled
bulk metallicities of Jupiter and Saturn. All eleven are indeed
metal-enriched, by factors ranging from 2--47 relative to the Sun,
and 2--20 relative to their host stars.
One caveat behind these results is that models of hot Jupiter
interiors are subject to the uncertainty over the extra source of
internal heat responsible for their unexpectedly large
radii (see, e.g., \citealt{batyginstevenson10}, who also describe a promising
solution). To inflate planetary radii, \citet{guillotetal06}
included in each hot Jupiter model an additional
source of power equal to 0.5\% of the received stellar irradiation
\citep{guillotshowman02}. The bulk metallicities inferred from the models
depend on the details of this extra energy source. Modulo this caveat, every planet is enriched in metals by at least a factor of $\sim$2 above solar, which is consistent with our finding that forming planetesimals by
gravitational instability requires metal enrichments of this order.


\acknowledgments We thank Xue-Ning Bai, John Johnson, Eve Ostriker,
Jim Stone, and Neal Turner for discussions, and Tristan Guillot for
the data in Table \ref{tab:metal}. Xue-Ning Bai, Anders Johansen, Jim
Stone, and Andrew Youdin provided valuable feedback on a draft version
of this paper. We are grateful to Stuart Weidenschilling for an
insightful referee's report that put our work into 
better context. This research was supported by the National Science
Foundation, in part through TeraGrid resources provided by Purdue
University under grant number TG-AST090079. A.T.L. acknowledges
support from an NSF Graduate Fellowship.


\appendix


\begin{deluxetable}{ccccccc}[h]
\tablecaption{Metallicities of Extrasolar Planets \citep{guillotetal06} and Solar System Gas Giants \citep{guillot05}.}
\tabletypesize{\small}
\tablewidth{0pt}
\tablehead{
\multicolumn{7}{c}{}\\
Name & 	$M_{\rm planet}$ & 	$M_{\rm Z}$$^a$ & 	$Z_{\rm planet}$ & 	$Z_{\rm planet}/Z_{\sun}$$^b$ & 	[Fe/H]$_{\ast}$ & 	$Z_{\rm planet}/Z_{\ast}$ \\ &  ($M_\earth$)     &  ($M_\earth$)  & ($M_{\rm Z}/M_{\rm planet}$) &  & &  \\ }
\startdata
HD209458 	& 210 	& 20 	& 0.095 	& 6.35 	& 0.02 & 6.06 \\
OGLE-TR-56	&	394	& 120	& 0.304 & 20.3 & 0.25 & 11.418 \\
OGLE-TR-113 &	429 & 70 & 0.163 & 10.9 & 0.15 & 7.7 \\
OGLE-TR-132 &  350 & 105 & 0.3 & 20 & 0.37 & 8.531 \\
OGLE-TR-111 & 168 & 50 & 0.297 & 19.84 & 0.19 & 12.81 \\
OGLE-TR-10   & 200 & 10 & 0.05 & 3.33 & 0.28 & 1.75 \\
TrES-1           & 238 & 50 & 0.21 & 14.0 & 0.06 & 12.2 \\
HD149026       & 114 & 80 & 0.70 & 46.78 & 0.36 & 20.42 \\
HD189733       & 365 & 30 & 0.082 & 5.479 & -0.03 & 5.87 \\ \hline
Jupiter            	&  318      & 10--42  	&0.03--0.13   	& 2.0--8.8 	& 0    		& 2.0--8.8    \\
Saturn            	&   95.2    	& 15--30 	&0.16--0.32     	& 11--21     	&  0        		&  11--21      \\
\enddata
\tablenotetext{a}{The metal content for each listed extrasolar planet was derived from a model of a planetary interior that includes an additional energy source
at the planet's center whose power equals 0.5\% of the incident stellar luminosity.}
\tablenotetext{b}{The solar metallicity $Z_\sun$ is taken to be 0.015 \citep{lodders03}. \\ \ \\}
\label{tab:metal}
\end{deluxetable}

\section{Background Disk Model}\label{app:background}
For numerical estimates in this paper, we adopt the standard
disk model derived in the review by \citet{chiangyoudin10}. The disk
has surface densities
\begin{eqnarray}
\label{eq_sigmag}
\Sigma_{\rm g} = 2200 \,F \left( \frac{r}{\rm AU} \right)^{-3/2} \gm \cm^{-2} \\
\label{eq_sigmad}
\Sigma_{\rm d} = 33 \,F\, \Zr \left( \frac{r}{\rm AU} \right)^{-3/2} \gm \cm^{-2}
\end{eqnarray}
in gas (g) and dust (d). The dimensionless parameters $F$ and $\Zr \equiv (\Sigma/\Sigmag)/0.015$,
typically of order unity, describe how much total mass the disk has
relative to the minimum-mass solar nebula, and how metal-rich the disk
is compared with a gas of solar abundances, respectively.  The
minimum-mass solar nebula ($F=1$, $\Zr=1$) uses a condensate mass
fraction for solar abundances of $\Sigmad/\Sigmag = 0.015$
\citep{lodders03}.  Values of $\Zr > 1$ correspond to supersolar
metallicities $\Sigmad/\Sigmag > 0.015$.
 Integrated to $r = 100\AU$,
equation (\ref{eq_sigmag}) yields a total disk mass of $0.03 F M_{\odot}$.

At the disk midplane, the gas temperature, scale height, and density are
given by
\begin{eqnarray}
\label{eq_T}
T &=& 120 \left( \frac{r}{\rm AU} \right)^{-3/7} \K \\
\label{eq_h}
H_{\rm g} &=& 0.022 
r \left( \frac{r}{\rm AU} \right)^{2/7} \\
\label{eq_rho}
\rho_{\rm g0} &=& 2.7  
\times 10^{-9} F \left( \frac{r}{\rm AU} \right)^{-39/14} \gm \cm^{-3} \,.
\end{eqnarray}
These are adapted from \citet{chianggoldreich97}, adjusted for a disk obeying
(\ref{eq_sigmag})--(\ref{eq_sigmad}), orbiting a pre-main-sequence
star of mass $M_{\ast} = 1 M_{\odot}$, radius $R_{\ast} = 1.7
R_{\odot}$, and temperature $T_{\ast} = 4350 \K$.

\section{The Super-Linear Relation Between Midplane Dust-to-Gas Ratio $\mu_0$ and Bulk Metallicity $\Sigmad/\Sigmag$}\label{app:superlinear}

We derive $\mu_0$ as a function of $\Sigmad/\Sigmag$ under the
assumption of a constant $Ri$. Some evidence supporting a constant
$Ri$ was found in our simulations (\S\ref{ssec:howconstant}). The density
profile for constant $Ri$ is used in a number of papers
(\citealt{sekiya98}; \citealt{youdinshu02}; Paper I) and we begin by
repeating the result, neglecting self-gravity as we have
throughout our paper. The dust-to-gas ratio is given by
\begin{equation}\label{eqn:b1}
\mu(z) = \left[\frac{1}{1/(1+\mu_0)^2 + (z/z_{\rm d})^2}\right]^{1/2} - 1
\end{equation}
where 
\begin{equation}
\label{eqn:dustheight}
z_{\rm d} \equiv \frac{Ri^{1/2}\,\vmax}{\OmegaK} 
\end{equation}
is a characteristic dust height and $\vmax = \eta \OmegaK r$ (see
equations \ref{eqn:vphi} and \ref{eqn:eta}) is a constant equal to the
difference in azimuthal velocity between a strictly Keplerian flow and
dust-free gas. The dust density drops to zero at
\begin{equation}
\label{eqn:dustmax}
z = \pm z_{\rm max} = \pm \frac{\sqrt{\mu_0(2+\mu_0)}}{1+\mu_0} z_{\rm d} \,.
\end{equation}

A comment on equation (\ref{eqn:b1}), in the limit that $\mu_0 \gg 1$:
except where $\mu$ is nearly constant near $z \ll z_{\rm max}/\mu_0$ and
where it falls to zero near $z = \zmax$, the shape of $\mu(z)$ is that
of $1/z$. This form follows simply from the constancy of $Ri$.
Because the numerator of $Ri$ is approximately constant with $\mu$
(\S\ref{ssec:superlinear}), the denominator must be as well: $\partial
v_{\phi}/\partial z \sim (\vmax/\mu) / z \sim$ constant, which implies
$\mu \propto 1/z$.  From this we can deduce
the super-linear trend between $\mu_0$
and $\Sigmad/\Sigmag$ as follows. The integral
of $\mu$ with respect to $z$ is proportional to the total surface
density of dust $\Sigmad$. Because $\mu \propto 1/z$, flattening
off as $z$ decreases below $z_{\rm max}/\mu_0$, this integral
varies as $\log \mu_0$. Then $\mu_0 \propto \exp \Sigmad$, crudely.

\begin{figure}[h]
\includegraphics[scale=0.90]{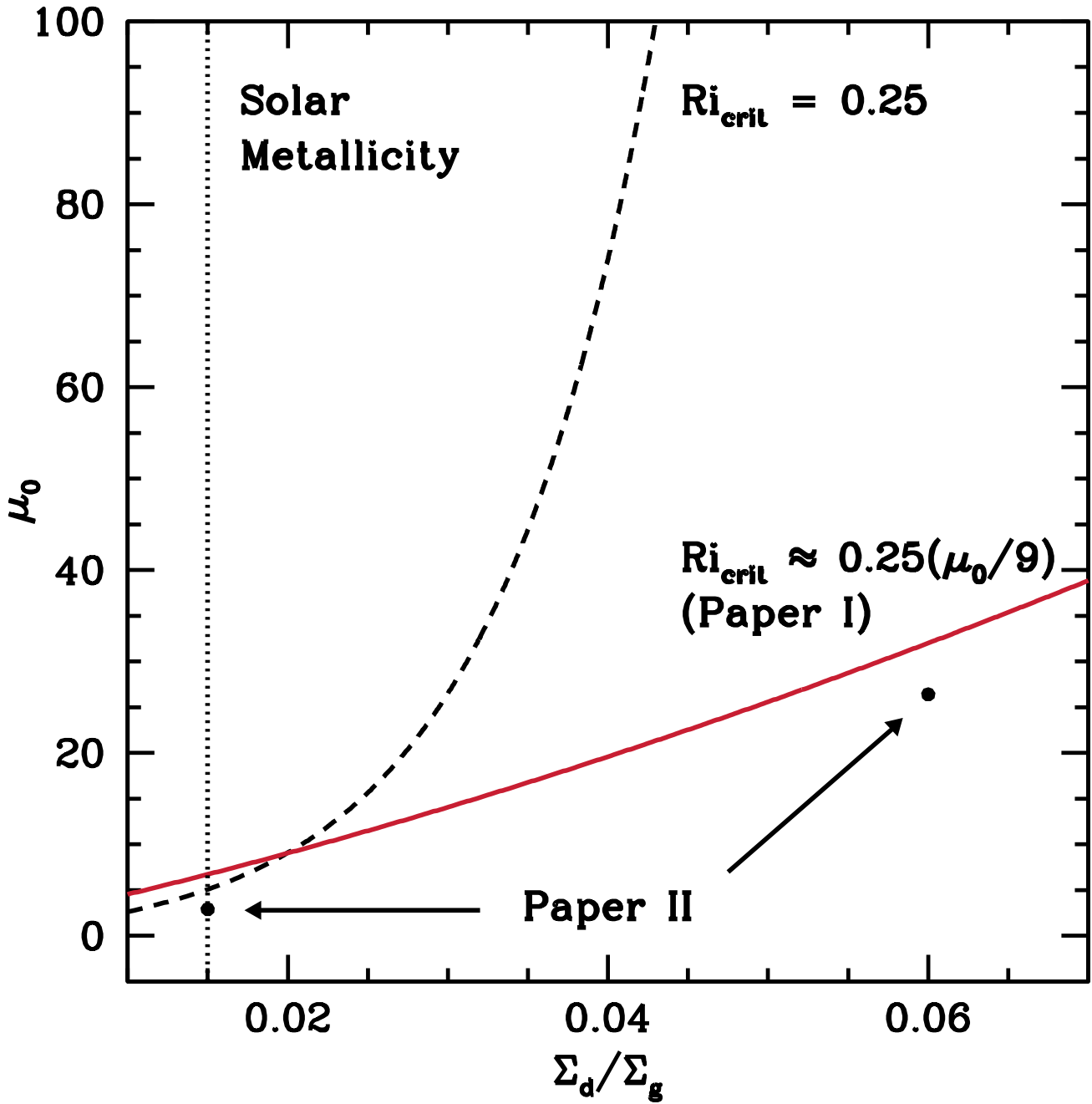}
\caption{Super-linear trend between the midplane dust-to-gas ratio
  $\mu_0$ and height-integrated metallicity $\Sigmad/\Sigmag$ for dust
  profiles characterized by a spatially constant Richardson number
  $Ri_{\rm crit}$. Equation \ref{eqn:superlinear} is plotted twice:
  the dashed curve uses $Ri_{\rm crit}=0.25$, whereas the solid curve
  varies $Ri_{\rm crit}$ according to the relation found in Paper I:
  $Ri_{\rm crit} \approx 0.25 (\mu_0/9)^{1.0}$ (see Figure 5 of Paper
  I).  Both curves fix $v_{\rm max}/c_{\rm s} = 0.025$. The maximum
  values of $\mu_0$ achieved in this paper are plotted as
  points. These data follow the variable $Ri_{\rm crit}(\mu_0)$ curve
  more closely than the constant $Ri_{\rm crit}$ curve, corroborating
  the evidence we found in \S\ref{ssec:howconstant} that $Ri_{\rm crit}$
  is spatially constant but varies with $\mu_0$ (equivalently
  $\Sigmad/\Sigmag$).}
\label{fig:superlin}
\end{figure}

More formally, we have
\begin{equation}
\Sigmad = 2 \int_0^{\zmax} \rhod\, dz = 2\rho_{\rm g0} \int_0^{\zmax} \mu\, dz
\end{equation}
where $\rho_{\rm g0}$ is the midplane gas density, assumed constant
because $\zmax \ll \Hg$. The gas density profile always well
approximates a Gaussian (see footnote \ref{foot:muchado}), from which
it follows that $\Sigmag \approx \sqrt{2\pi} \rho_{\rm g0} \Hg$.  Then
\begin{equation}\label{eqn:integral}
  \frac{\Sigmad}{\Sigmag} = \sqrt{ \frac{2}{\pi} } \frac{1}{\Hg} \int_0^{\zmax} \mu \, dz \,.
\end{equation}
Inserting (\ref{eqn:b1}) into (\ref{eqn:integral}) we have
\begin{equation}\label{eqn:superlinear}
\sqrt{ \frac{\pi}{2} } \frac{H_{\rm g}}{z_{\rm d}}  \frac{\Sigmad}{\Sigmag} = \log [1 + \mu_0 + \mu_0^{1/2}(2+\mu_0)^{1/2}] - \frac{\mu_0^{1/2}(2+\mu_0)^{1/2}}{(1+\mu_0)} \,.
\end{equation}
In the limit $\mu_0 \gg 1$, the exponential dependence of $\mu_0$
on $\Sigmad/\Sigmag$ is evident.  Equation (\ref{eqn:superlinear}) is
plotted in Figure \ref{fig:superlin}, with $Ri=0.25$ and $v_{\rm
  max}/c_{\rm s} = 0.025$. Overlaid is the same equation but with
varying $Ri = Ri_{\rm crit} \approx 0.25 (\mu_0/9)$, the relation we
found in Paper I (see Figure 5 of that paper). The two data points
representing the maximum $\mu_0$ achieved in this paper are also
plotted. The data track the variable $Ri_{\rm crit}(\mu_0)$ curve much
better than the constant $Ri$ curve.

Finally note that $H_{\rm g}/z_{\rm d} \propto \cs/\vmax$ enters into equation (\ref{eqn:superlinear})
the same way that $\Sigmad/\Sigmag$ does. Thus $\mu_0$ increases super-linearly
with $\cs/\vmax$ as well. This result leads us to suspect that our numerical
results for $\mu_0$ (2.9, 26.4) depend sensitively on our choice
for $\vmax/\cs = 0.025$. In this paper we did not run simulations with
different $\vmax/\cs$ and so did not test this suspicion.

\end{document}